\documentclass[sn-mathphys]{sn-jnl}

\jyear{2022}%

\theoremstyle{thmstyleone}%
\theoremstyle{thmstyletwo}%

\theoremstyle{thmstylethree}%

\begin{document}

\title[A Spectrum of Galactic Rotation Curves]{Dark Energy and Extending the Geodesic Equations of Motion: A Spectrum of Galactic Rotation Curves}


\author*[1,2]{\fnm{Achilles D.} \sur{Speliotopoulos}}\email{ads@berkeley.edu}

\affil*[1]{\orgdiv{Department of Physics}, \orgname{University of California}, \orgaddress{\street{}
    \city{Berkeley}, \postcode{94720}, \state{CA}, \country{USA}}}

\affil[2]{\orgdiv{Division of Physical Sciences and Engineering}, \orgname{Diablo Valley College}, \orgaddress{\street{321 Golf Club Road}, \city{Pleasant Hill}, \postcode{94523}, \state{CA}, \country{USA}}}


\abstract{A recently proposed extension of the geodesic equations of
  motion, where the worldline traced by a test particle now depends on
  the scalar curvature, is used to study the formation of galaxies and
  galactic rotation curves. This extension is applied to the motion of a
  fluid in a spherical geometry, resulting in a set of
  evolution equations for the fluid in the nonrelativistic and weak
  gravity limits. Focusing on the stationary solutions of these
  equations and choosing a specific class of angular
  momenta for the fluid in this limit, we show that 
  dynamics under this extension can result in the formation
  of galaxies with rotational velocity curves (RVC) that are consistent with
  the Universal Rotation Curve (URC), and through previous work on the
  URC, the observed rotational velocity profiles of 1100 spiral
  galaxies. In particular, a spectrum of RVCs can form 
  under this extension, and we find that the two extreme velocity curves
  predicted by it brackets the ensemble of the URCs constructed from these 1100
  velocity profiles. We also find that the asymptotic behavior of the URC
  is consistent with that of the most probable asymptotic behavior of the
  RVCs predicted by the extension. A stability
  analysis of these stationary solutions is also done, and we find
  them to be stable in the galactic disk, while in the galactic hub
  they are stable if the period of oscillations of perturbations is
  longer than $0.91_{\pm0.31}$ to $1.58_{\pm 0.46}$ billion years. 
}




\maketitle

\section{Introduction}

With the discovery of dark energy $\Lambda_{DE}=
(7.21_{-0.84}^{+0.83})\times 10^{-30}$ g/$\hbox{cm}^3$  
\cite{Ries1998} - \cite{WMAP}, comes a universal length scale
$\lambda_{DE} = c/\left(\Lambda_{DE}G\right)^{1/2}=14010_{820}^{800}$
Mpc, for the universe that allows for extensions of the geodesic
equations of motion (GEOM). However, to be physically viable these
extensions must overcome high hurdles. As outlined in \cite{ADS2010a},
these hurdles include the following: Ensuring that the equivalence
principle is preserved; this principle is one of the underlying principles upon
which general relativity is founded, and has been experimentally
verified. Requiring that the equations of motion for massless test
particles are not affected; all astronomical observations\textemdash
in particular, those with which the rotational velocity profiles of
spiral galaxies are determined\textemdash are based on the motion of
photons. Demonstrating that the extension is not prevented by attempts at
showing the GEOM is the unique consequence of Einstein's field
equations \cite{Ein1938}-\cite{Ehl2003}; such proofs limit the
structure of possible extensions. Finally,
ensuring that effects which could have already been measured in 
terrestrial experiments, or observed in the motion of bodies in the
solar system, are not produced; such extensions would have 
been automatically ruled out by experiment.

In \cite{ADS2010a} we proposed an extension of GEOM, called the
\textit{extended GEOM}, that satisfies these conditions. It was
constructed using the dimensionless parameter $c^2 R/\lambda_{DE}G$,
where $R$ is the Ricci scalar, and replacing the mass $m$ of the test particle by
$m\mathfrak{R}[c^2R/\Lambda_{DE}G]$ in the Lagrangian for a test
particle in general relativity. By doing so we have changed the
response of the test particles to the geometry of spacetime; the worldlines of
massive test particles now depend on the local scalar curvature of the
spacetime. Importantly, Einstein's field equations are not changed,
and thus the geometry of spacetime is still determined by them. The
degree by which the worldline is changed is 
determined by $\mathfrak{R}$, which is taken to be a non-linear
function of $c^2R/\Lambda_{DE}G$ 
with the nonlinearity modulated by a single parameter, the power-law
exponent $\alpha_{{}_\Lambda}$. This exponent determines the
asymptotic behavior of $\mathfrak{R}$ for large arguments. A strict
lower bound, $\alpha_{\Lambda \hbox{\scriptsize{Bound}}}$, for
$\alpha_{\Lambda}$ was determined with the range of possible values for
$\alpha_{\Lambda\hbox{\scriptsize{Bound}}}$ established in
\cite{ADS2010a} by requiring that signatures of the GEOM must not
have already been seen in terrestrial experiments. With reasonable
choices for experimentally measurable parameters, we found that
$1.28\le\alpha_{\Lambda \hbox{\scriptsize{Bound}}} \le 1.58$

Given the scale of $\lambda_{DE}$, it is only at galactic length
scales or longer where the impact of the extended 
GEOM is expected to be seen, and in \cite{ADS2010b} we applied this
extension to the analysis of the motion of bodies at these
scales. Using a spherical model for galaxies, we calculated the
density profile of a stationary  galaxy given the radius
$r_H^{*}=11.82_{\pm 0.30}$ kpc of a typical 
galactic hub and the velocity $v_H^{*}=172.1_{\pm 1.6}$ km/s of a
typical rotational velocity curve (RVC) at this radius
\cite{ADS2010b}. This $r_H^{*}$ and $v_H^{*}$ were determined from the
observed motion of stars in 1,393 spiral galaxies \cite{Cour} -
\cite{Rubin1985}. The density profile for the model galaxy was
determined using the extended GEOM and the following model of the RVC
of the galaxy, 
  \[v^{\hbox{\scriptsize ideal}}(r) = \left\{ 
\begin{array}{l l}
  v_H r/r_H & \quad \mbox{for $r \le r_H$}\\
  v_H & \quad \mbox{for $r\ge r_H$, }\\ 
\end{array} 
\right\},
\]
where $v_H$ is the asymptotic velocity of the curve. The power-law
exponent was set to $\alpha_{{}_\Lambda} = 
1.56_{\pm 0.10}$ using the Hubble length and the density of this model
galaxy (the details of this analysis can be found in
\cite{ADS2010b}); this value is within the bounds for
$\alpha_{\Lambda}$ found 
in \cite{ADS2010a}. The radius $R_{200}$ for this density profile was
calculated to be $206_{\pm 53}$ kpc, in reasonable agreement with
observations. Importantly, $\sigma_8$ was also calculated, and was
found to be $0.73_{\pm 0.12}$, which is within experimental error of
both the WMAP value of $0.71_{-0.048}^{+0.049}$ \cite{WMAP}, and the
PLANCK value of $0.81_{\pm 0.006}$ \cite{PLANCK}. 

In \cite{ADS2010a} and \cite{ADS2010b} the focus was on using the
extended GEOM to determine the properties of a stationary galaxy
that has already been formed. However, if the values of $R_{200}$ and $\sigma_8$
measured are due to the extended GEOM, then the formation of galaxies
must be describable, and the possible RVCs for these galaxies
predictable, within this extension. Yet, given the drastic difference
between galactic length scale (on the order of tens of kiloparsecs)
and the length scales at which $R_{200}$ and $\sigma_8$ are relevant
(on the order of a few hundred kiloparsecs and a few 
megaparsecs, respectively), the results
of our previous paper speaks little about the formation of
galaxies. Indeed, since a specific velocity curve
$v^{\hbox{\scriptsize{ideal}}}(r)$ was used to begin with, it
certainly cannot predict the RVCs of them. The purpose of this
paper is address this lack, and to begin fulfilling these
expectations. In particular, our goal here is to establish the range
of possible asymptotic behaviors of the RVCs that are allowed by the
extended GEOM, and to compare these predictions with observations.  

In \cite{ADS2010a} we showed that the energy-momentum tensor
$T_{\mu\nu}$ for a collection of massive particles that can be treated
as a fluid with density $\rho$,
and fluid velocity $u^\mu$ reduces in the nonrelativistic
limit to $T_{\mu\nu} \approx \rho u_\mu u_\nu$ even when elements of the
fluid evolve under the extended GEOM. Applying that result here, to 
a spherically symmetric distribution of particles rotating about a
single rotational axis, we obtain a set of evolution equations
\textbf{\textit{Evol}} for the density $\rho(t,r)$; the fluid velocity along
the radial direction $u^r(t,r)$; the gravitational potential
$\Phi(t,r)$; and the (specific) angular momentum $L(t,r)=rv(t,r)$, where
$v(t,r)$ is the rotational velocity of the fluid about the rotational
axis. Importantly, as our extension of the GEOM involves replacing the
mass $m$ of a test particle by $m\mathfrak{R}[c^2R/\Lambda_{DE}G]$,
and as this replacement is the same for all particles irrespective of
its nature (as such the extended GEOM obeys the weak equivalence
principle), the extended GEOM\textemdash and throught it the
\textbf{\textit{Evol}}\textemdash does not differentiate between
baryonic and  dark matter; the density $\rho$ of the fluid is the total
density of matter in the model galaxy. We have shown below that both
the mass and the angular momentum of the system are conserved under
this evolution. The types of galaxies that can form, and the RVCs that
they can have, under the extended GEOM would then be determined by the
solution of \textbf{\textit{Evol}} for some initial distribution of
mass and velocities. These evolution equations are extremely nonlinear,
however, and it is doubtful that any direct attempt at solving them
will yield much of use. We have taken a different approach instead. 

If a choice of the initial distribution of mass and velocities results
in the formation of a galaxy under the extended GEOM, then the
resultant distribution of mass and velocities must result in stationary
solutions\textemdash denoted by $\rho_{\infty}(r), L_{\infty},
\Phi_{\infty},$ and $u^r_{\infty}$\textemdash of
$\hbox{\textbf{\textit{Evol}}}$. Focusing further on galaxies where the motion
of matter traces out circular orbits and $\hbox{\textbf{\textit{Evol}}}$ reduces
to Eq.~$(11)$ of \cite{ADS2010b}, a single second-order, nonlinear,
inhomogeneous differential equation for the stationary density
$\rho_{\infty}(r)$ of the galaxy with the inhomogeneous term given by the
angular momentum $L_{\infty}(r)$ of the fluid in this
limit. Importantly, solutions $\rho_{\infty}(r)$ of this differential
equation minimizes a stationary action ${S}_{\infty}$. The dependence
of the structure of the galaxy on $L_{\infty}(r)$\textemdash and
given that the total angular
momentum is conserved, on the initial distribution of 
angular momentum $L(0,r)$ of the fluid\textemdash underscores the
important role that angular momentum plays in the formation of 
galaxies even under the extended GEOM. While it is in principle possible
to choose an initial $L(0,r)$, and then use it to determine whether a
galaxy can form under the extended GEOM with this choice, and if 
it can, whether such a galaxy has a RVC that agrees with observations,
doing so would mean evolving $L(0,r)$ in time to $L_{\infty}(r)$ using
$\hbox{\textbf{\textit{Evol}}}$. This likely is also intractable
analytically.  

Since a choice of angular momentum for the fluid must be made, we make
this choice at the stationary limit instead of at the fluid's initial
state. Using the observed properties of galaxies, we focus on a 
class of stationary angular momentum given by the RVC 
\begin{equation}
  v_{\infty}(r) =  \left[\frac{(1+p/q)x^{2q}}{x^{2(q+p)}+p/q}\right]^{1/2}v_H^{*},
  \label{e1}
\end{equation}
where $x=r/r_H^{*}$. Here, $q$ and $p$ are parameters that give the
asymptotic behavior of $v_{\infty}(r)$ in the $x\ll 1$ and $x\gg1$
limits, respectively, with the subscript denoting that we are in the
stationary limit. With this choice we are able to
determine whether galaxies can form under the extended GEOM, and
will be able to predict their RVC. The choice
itself depends only on four parameters, each of which have good physical
interpretation, and each of which can either be determined (for
$v_H^{*}$ and $r_H^{*}$)\textemdash and thus used as inputs in the
analysis\textemdash through observations, or compared (for $q$ and
$p$) to them. This choice is a natural generalization of
$v^{\hbox{\scriptsize ideal}}(r)$ that is also a smooth function of $r$, a condition
that is important for both physical and mathematical
reasons. Importantly, with two free parameters and with
$L_{\infty}(r)=rv_{\infty}(r)$, $v_{\infty}(r)$  can model a variety of possible angular momenta  
for the fluid in the stationary limit, and thus has the potential to model a variety 
of possible angular momentum $L(0,r)$ at the system's initial
state. It thereby defines a class of rotational velocity
profiles, one for each given $q$ and $p$, and importantly, the
predicted values of the $q$ and $p$ obtained through the extended GEOM
can be directly compared to observations.  

To determine the values of $q$ and $p$ that will result in a
stationary galaxy under the extended GEOM, we make use of
$S_{\infty}$. Each choice of $q$ and $p$ results in a
$L_{\infty}(r; q, p)$, which in turn results in 
a solution $\rho_{\infty}(r;q,p)$ of $\hbox{\textbf{\textit{Evol}}}$ in the
stationary limit. Such a choice for $L_{\infty}(r; q, p)$ need not, in
general, lead to a $\rho_{\infty}(r;q,p)$ that minimizes 
$S_{\infty}$, however. Thus, not all choices of $q$ and $p$ will
result in the formation of a galaxy under the GEOM. To determine the
values of $q$ and $p$ that do, we evaluate $S_{\infty}\vert_{(\rho_{\infty};
  L_{\infty})}$ at this $\rho_{\infty}(r;q,p)$ and
$L_{\infty}(r; q, p)$; the resultant action then depends on the
parameters $q$ and $p$. Minimization of this action with respect to
these parameters then gives the values of $q$ and $p$ that, when used
in $L_{\infty}(r; q, p)$, gives the $\rho_{\infty}(r;q,p)$ that
does minimize $S_{\infty}$. It is for these values of
$q$ and $p$ that the extended GEOM would predict a galaxy can form. (This
approach in determining $q$ and $p$ follows a minimization 
principle that is similar to the least squares and Rayleigh-Ritz
variational methods for solving differential equations
\cite{Zwil1989}. Like those methods, the resultant $L_{\infty}(r)$
and $\rho_{\infty}(r)$ obtained are an approximation of the solution of
\textbf{\textit{Evol}} in the stationary limit.) If no such $q$ and
$p$'s can be found, then this choice of $v_{\infty}(x)$ for a
class of possible rotational velocity profiles of galaxies is too
limited. Galaxies with a RVC given by $v_{\infty}(x)$\textemdash and
likely even those approximated by it\textemdash cannot be formed under the
extended GEOM.   

At the end of this analysis, we find that the action
$S_{\infty}$ does not have one local minimum\textemdash or
even a discrete number of local minima\textemdash for a distinct pair
of $(q,p)$. Rather, for each choice of $q$ between $0.010$ and
$0.336$ there is a $p$ between $0.348$ and $0.480_{\pm
  0.02}$ that minimizes $S_{\infty}$. The asymptotic
behavior of the RVC in the galactic hub is thus connected with the
asymptotic behavior of the RVC outside of it. This dependence between the
two parameters is expected. A single galaxy is formed from a single
fluid, and during its formation, fluid elements in
one region will interact with the fluid elements in
other regions of it. To have the structure of the galaxy
inside of the galactic hub be independent of the
structure of the galaxy outside of it is physically unreasonable.

That the extended GEOM predicts the formation of a variety of
galaxies, each with a different density profile, is a result that is
certainly consistent with observations. That the predicted RVCs for these
galaxies are different is consistent with both observations and
the \textit{Universal Rotation Curve} (URC) proposed by Persic et.~al.

In \cite{Pers1996} Persic et.~al.~analyzed a homogeneous
sample of 1100 RVCs of spiral galaxies, and found that only one
global parameter\textemdash the luminosity of the galaxy\textemdash
determines the profile of the RVC observed. To describe this
dependency they proposed the universal 
rotation curve $V_{URC}$, which gives the velocity profile of
any galaxy given its luminosity. Salucci et.~al.~further refined the
URC model in \cite{Salu2007}, and applied it to
the RVC of spiral galaxies; this refinement  
was then applied to dwarf spheroidal galaxies and low surface brightness
galaxies in \cite{Salu2012} and \cite{Paol2019}, respectively. One of
the main results of \cite{Salu2007} is shown in Fig.~4 of that paper,
where the authors plotted an ensemble of URCs, each with a
different virial mass. To show the similarity between the curves and
to compare these curves to $V_{NFW}$, the RVC obtained
from $N$-body simulations of Lambda cold dark matter \cite{Nava1996},
all of the curves were rescaled and normalized to agree at
the viral radius. We find that the RVCs predicted here by the extended
GEOM agrees well with the curves shown in this figure.  

The spectrum of RVCs predicted by the extended
GEOM ranges from $(q,p) = (0.010, 0.048_{\pm 0.020})$ to
$(0.336, 0.387_{\pm0.090})$, with the median curve given by
$(0.172, 0.349_{\pm 0.010})$. When $v_{\infty}(x)$ is rescaled and normalized
to fit the scale used in Fig.~4 of \cite{Salu2007}, we find that the
ensemble of curves from $V_{URC}$ is bracketed below by the $(0.010,
0.048_{\pm 0.020})$ curve and above by the $(0.336, 0.387_{\pm0.090})$
curve; the median curve $(0.172, 0.349_{\pm 0.010})$ lies in the
middle of the ensemble of URC curves, and is surprisingly close to the 
$V_{NFW}$ curve. In addition, we find that the most probable
asymptotic behavior in the large $x$ limit for the RVCs predicted by the
extended GEOM has a $p=0.348$, in good agreement with the profile
for $V_{NFW}$, which has an asymptotic power-law exponent of
$0.33+\epsilon_{NFW}$ with $\epsilon_{NFW}<0.1$ \cite{Salu2007}.

While the minimization of $S_{\infty}$ does show that stationary
galaxies with $L_{\infty}(r)$ can form and does predict the RVCs for
these galaxies, this analysis cannot determine whether the galaxies
predicted are stable under perturbations. To address this lack, we have also
completed a stability analysis of the predicted galaxies by
perturbing about stationary solutions of \textbf{\textit{Evol}}. This
results in a second-order, partial differential equation for first-order 
perturbations of the stationary radial velocity. We find that the
region outside of the galactic hub is very rigid; small perturbations
in the radial velocity remain small no matter the frequency of the
perturbation. Within the galactic hub, on the other 
hand, we find that when the frequency of the perturbation is smaller
than $0.267_{\pm 0.076}$ to $0.47_{\pm0.16}$ times the maximum angular
velocity of the hub (corresponding to a period of $0.91_{\pm0.31}$ to
$1.58_{\pm 0.46}$ billion years) then perturbations in the radial
velocity remains small. If, however, it is larger than this range of angular
velocities then within the galactic hub small perturbations can increase
exponentially with radius. 

The rest of the paper is organized as follows. In
\textbf{Sec \ref{Evol}} the focus is on the evolution of fluids under
the extended GEOM. The spherical 
model of the fluid used in this paper is presented. Difficulties in
applying \textbf{\textit{Evol}} to the formation of galaxies is
pointed out, and an alternative approach using the stationary limit of
\textbf{\textit{Evol}} is proposed. Details of this approach
is given in \textbf{Sec \ref{Stat}}, and the important role that the
angular momentum plays is shown. A specific form for
$L_{\infty}(r)$ is proposed. Approximate solutions to the stationary limit of
\textbf{\textit{Evol}} are found in \textbf{Sec \ref{Solu}} using
techniques from boundary layer theory. It is then found in \textbf{Sec
  \ref{Spec}} that a spectrum of RVCs is formed under the extended
GEOM, and the range of this spectrum is determined. Comparisons with
the URC are then made. A stability analysis of the stationary solutions is
presented in \textbf{Sec \ref{Stab}}, and concluding remarks can be
found in \textbf{Sec \ref{End}}.  

\section{Evolution under the Extended GEOM}
\label{Evol}

In this section we focus on the time evolution of fluids under the extended
GEOM, and the use of these evolution equations in determining the formation of
galaxies. We begin with a brief review of the extended GEOM as
applied to individual test particles. These equations of motion are
then applied to the motion of the collection of these particles that form a
fluid using the energy-momentum tensor for the fluid in the 
nonrelativistic and weak-gravity limits. A spherical model of a galaxy is
then presented, and the \textbf{\textit{Evol}} is obtained. The
difficulties in using \textbf{\textit{Evol}} to determine the structure of
galaxies are pointed out, and an alternate approach using the
stationary limit of \textbf{\textit{Evol}} is proposed.

\subsection{A Review of the Extended GEOM for Test Particles}

As WMAP measured the pressure to energy density ratio for Dark
Energy to be $-0.967^{+0.073}_{-0.072}$ \cite{WMAP}\textemdash within experimental
error of the ratio expected for the cosmological constant\textemdash
following \cite{ADS2010b} we identify Dark Energy with the
cosmological constant, and required only that $\Lambda_{DE}$ changes
so slowly that it can be considered a constant. Einstein's field
equations are then 
\begin{equation}
  R_{\mu\nu} - \frac{1}{2}g_{\mu\nu}R +
  \frac{\Lambda_{DE}G}{c^2} g_{\mu\nu}= - \frac{8\pi G}{c^4} T_{\mu\nu},
\label{e2}
\end{equation}
where $T_{\mu\nu}$ is the energy-momentum tensor for matter,
$R_{\mu\nu}$ is the Ricci tensor, Greek indices run from $0$ to $3$,
Latin indices run from $1$ to $3$, and the signature
of $g_{\mu\nu}$ is $(1,-1,-1,-1)$. Here, we have followed \cite{Wald1984} and taken,
\begin{equation}
  R_{\mu\nu,\alpha}^{\>\>\>\>\quad\beta}=\partial_\nu\Gamma_{\mu\alpha}^{\beta}-
    \partial_\mu\Gamma_{\nu\alpha}^{\beta}+
      \Gamma^{\lambda}_{\mu\alpha}\Gamma_{\lambda\nu}^\beta
      -\Gamma^{\lambda}_{\nu\alpha}\Gamma_{\lambda\mu}^\beta,
\end{equation}
while
\begin{equation}
\Gamma^\alpha_{\mu\nu} = \frac{1}{2}g^{\alpha\beta}\left(\partial_\mu
g_{\nu\beta} + \partial_\nu g_{\beta\mu}-\partial_\beta g_{\mu\nu}\right).
\label{e-G}
\end{equation}

The extended GEOM for a test particle with mass $m$ is obtained from
the Lagrangian
\begin{equation}
\mathcal{L}_{\hbox{\scriptsize{Ext}}} \equiv
mc \mathfrak{R}[c^2R/\Lambda_{DE}G]
\left(g_{\mu\nu}\dot{x}^\mu\dot{x}^\nu\right)^{\frac{1}{2}},
\label{e-L}
\end{equation}
where \textit{for this section only} $x$ is the four-vector,
$x^\mu=(x^0, x^1, x^2, x^3)$. In \cite{ADS2010a} we argued for
\begin{equation}
  \mathfrak{R}(c^2R/\Lambda_{DE}G) =
  \left[1+\mathfrak{D}(c^2R/\Lambda_{DE}G)\right]^{1/2},
  \label{e-R}
\end{equation}
where  
\begin{equation}
\mathfrak{D}(c^2R/\Lambda_{DE}G) = \chi(\alpha_{{}_\Lambda})\int_{c^2R/\Lambda_{DE}G}^\infty
\frac{ds}{1+s^{1+{\alpha_{{}_\Lambda}}}},
\label{e-D}
\end{equation}
while 
\begin{equation}
\frac{1}{\chi(\alpha_{{}_\Lambda})}\equiv \int_0^\infty
\frac{ds}{1+s^{1+{\alpha_{{}_\Lambda}}}}
=\frac{\sin\left[\pi/(1+\alpha_{\Lambda})\right]}{\pi/(1+\alpha_{\Lambda})}, 
\end{equation}
is chosen so that $D(0)=1$. Here, $\alpha_{\Lambda}$ is a constant, and is the
only free parameter in the theory.  To prevent the effects of the
extension from being already seen in terrestrial experiments, we
considered in \cite{ADS2010a} an experiment designed to look for
anomalous accelerations through the propagation of sound waves in a
gas of He${}^4$ atoms at $4$ K. Reasonable choices for experimental
parameters then gives the lower bound for $\alpha_{{}_\Lambda}$ to be
between $1.28$ (for $\Lambda_{DE} = 10^{-32}$ g/cm${}^3$) and 1.58
(for $\Lambda_{DE} = 10^{-29}$ g/cm${}^3$). 

From Eq.~$(\ref{e-L})$, the canonical momentum for the particle is
\begin{equation}
  p_\mu=mc\mathfrak{R}[c^2R/\Lambda_{DE}G]\frac{\dot{x}_\mu}{\left(\dot{x}^\mu\dot{x}_\mu\right)^{1/2}},
\end{equation}
leading to the constraint,
\begin{equation}
  p^2 = m^2 c^2 \left(\mathfrak{R}[c^2R/\Lambda_{DE}G]\right)^2,
\end{equation}
as expected.

As
\begin{equation}
  \frac{\partial\mathcal{L}_{\hbox{\scriptsize{Ext}}}}{\partial x^\mu}
  =\frac{mc}{\sqrt{\dot{x}^2}^{1/2}}\left(\frac{1}{2}
  \partial_\mu g_{\alpha\beta}\dot{x}^\alpha\dot{x}^\beta \mathfrak{R}+
      g_{\alpha\beta}\dot{x}^\alpha\dot{x}^\beta\frac{\partial \mathfrak{R}}{\partial x^\mu}\right),
\end{equation}
then with the parametization $\dot{x}^2=c^2$ the Euler-Lagrange equation gives
\begin{equation}
0=
\frac{d\>\>\>}{dt}\left(\mathfrak{R}g_{\mu\lambda}\dot{x}^\lambda\right)
- \left(\frac{1}{2}
  \partial_\mu g_{\alpha\beta}\dot{x}^\alpha\dot{x}^\beta \mathfrak{R}+
  c^2\partial_\mu \mathfrak{R}\right),
\end{equation}
or
\begin{equation}
0 =\mathfrak{R}\left[g_{\mu\lambda}\ddot{x}^\lambda
  +\partial_\nu g_{\mu\lambda}\dot{x}^\nu\dot{x}^\lambda-\frac{1}{2}
  \partial_\mu g_{\alpha\beta}\dot{x}^\alpha\dot{x}^\beta
  -\left(c^2\delta^\nu_\mu -\dot{x}_\mu\dot{x}^\nu\right)\partial_\nu\log\mathfrak{R}\right].
\end{equation}
It then follows from Eq.~$(\ref{e-G})$ that the extended GEOM for point particles is
\begin{equation}
\frac{D^2 x^\mu}{\partial t^2} = c^2 \left(g^{\mu\nu} - \frac{v^\mu
  v^\nu}{c^2}\right)
  \nabla_\nu\log\mathfrak{R}[c^2R/\Lambda_{DE}G].
\label{e3}
\end{equation}

It is important to note that we have not changed Einstein's field
equtions, and thus the geometry of spacetime is still given by the
solution of Eq.~$(\ref{e-L})$. What we have done by replacing $m$
with $m\mathfrak{R}[c^2R/\Lambda_{DE}G]$ in the Lagrangian for a test
particle in general relativity is to change the response of the motion
of test particles to the geometry of spacetime. As a consequence, the
worldline of the test particle is now given by the extended GEOM
Eq.~$(\ref{e3})$ and not the geodesic equations of motion.

Finally, in \cite{ADS2010b} we used Eq.~$(\ref{e3})$ to determine the
density profile of a galaxy with the velocity profile
$v^{\hbox{\scriptsize{ideal}}}(r)$ given in the introduction. This was
done by splitting the space around the model galaxy into three
regions. While the analysis in \cite{ADS2010b} for the first two
regions will change in this paper, the analysis in the third region
will not. Importantly, we found that in this third region the density of a
galaxy with a RVC given $v^{\hbox{\scriptsize{ideal}}}(r)$ by
decreases exponentially fast at distances greater than
$r_{II}=\sqrt{\chi/(1+4^{1+\alpha_{{}_\Lambda}})}\lambda_{DE}$ from
the center of the galaxy; the reader is referred to \cite{ADS2010b} 
for the details of this analysis. Since this decrease in density is not seen, the
maximum distance between galaxies is $2r_{II}$; setting this equal to
the Hubble length gives $\alpha_{\Lambda}=1.56_{\pm 0.10}$. 
  
\subsection{The Evolution of Fluids under the Extended GEOM}

We begin by considering a collection of particles in a region of space
that can be described as a fluid. The
distribution of such a fluid is given by its density $\rho(x)$, while
the four-velocity field for the fluid is given by the velocity field
$u^\mu(x)$. Then from Eq.~$(\ref{e3})$ the four-velocity of
each fluid element is given by the solution of the equation of motion, 
\begin{equation}
  u\cdot\nabla u^\mu = c^2
  \left(
  g^{\mu\nu}- \frac{u^\mu u^\nu}{c^2}
  \right)\nabla_\nu\log\mathfrak{R}.
    \label{e4}
\end{equation}

As we are interested in the formation of galaxies we work in the
nonrelativistic limit. In particular, in limit
$\rho c^2 \gg 3p$ we showed in \cite{ADS2010a} that by using the extended
GEOM Eq.~$(\ref{e3})$ the energy-momentum tensor for this fluid can
be approximated as $T_{\mu\nu}\approx\rho u_\mu u_\nu$, and the current density
$j_\mu\equiv T_{\mu\nu}u^\nu = \rho u_\mu$ is conserved: $\nabla\cdot j
\approx 0$. Next, WMAP and the Supernova Legacy Survey  put
$\Omega_K = -0.011_{\pm 0.012}$, and the spatial curvature is within
experimental error of vanishing. The universe is essentially spatially
flat. As the timescales and the lengthscales we are interested in are
much shorter than cosmological scales, we approximate the scale factor
in the Freeman-Lemaitre-Robertson-Walker metric to be a constant. We
are therefore working in the weark gravity limit, and can take the
metric to be to be $g_{\mu\nu} = \eta_{\mu\nu}+h_{\mu\nu}$.  Here, 
$\eta_{\mu\nu}$ is a flat background metric, and $h_{\mu\nu}$ is a
small perturbation of it that has only the one nonzero component:
$h_{00}=2\Phi/c^2$. We make this choice for $g_{\mu\nu}$ even though
with the $\Lambda_{DE}$ term in Eq.~$(\ref{e2})$ the spacetime will be
significantly different from Minkowski space at scales comparable to
$\lambda_{DE}$. At $14010^{800}_{820}$ Mpc,
$\lambda_{DE}$ is much larger then the length scales we are interested in,
however, and taking the background metric to be flat is a good
approximation.  

Writing the connection as $\Gamma^\alpha_{\mu\nu} =
{}_{{}_0}\Gamma^\alpha_{\mu\nu} +H^\alpha_{\mu\nu}$, then
\begin{equation}
  {}_{{}_0}\Gamma^\alpha_{\mu\nu}=\frac{1}{2} \eta^{\alpha\beta}
  \left(\partial_\mu\eta_{\nu\beta}+\partial_\nu\eta_{\beta\mu}-\partial_\beta\eta_{\mu\nu}\right),
\end{equation}
is the connection in the absence of matter and thus determined
solely by the coordinates used, while the contribution to
$\Gamma^\alpha_{\mu\nu}$ due to matter is
\begin{equation}
  H^\alpha_{\mu\nu} \equiv \frac{\partial_\mu\Phi}{c^2}  \eta^{\alpha
    0} \delta^0_\nu + \frac{\partial_\nu\Phi}{c^2} \eta^{\alpha
    0} \delta^0_\mu - \eta^{\alpha\beta}\frac{\partial_\beta\Phi}{c^2} \delta^{0}_\mu
      \delta^0_\nu.
\end{equation}
As expected, the 
non-vanishing components of ${}_{{}_0}\Gamma^\alpha_{\mu\nu}$ are
${}_{{}_0}\Gamma^k_{ij}$, while for $H^\alpha_{\mu\nu}$ they are
$H^0_{0i} = \partial_i\Phi/c^2$ and $H^i_{00} = -\eta^{ij}\partial_j\Phi/c^2$;
$H^0_{00}= \partial_t\Phi/c^3 \approx 0$ in the nonrelativistic limit. 

As $u^0\approx c$, both sides of the $\mu=0$ component of
Eq.~$(\ref{e4})$ is of order $u^i/c$, and are negligible in the
non-relativistic limit. The spatial components do survive, however,
and give
\begin{equation}
  \frac{\partial u^k}{\partial t} + u^i\partial_i u^k
  +u^iu^j{}_{{}_0}\Gamma^k_{ij} - \eta^{kl}\partial_l\left(\Phi +
  c^2\log\mathfrak{R}\right) + u^ku^i\partial_i \log\mathfrak{R}=0,
  \label{e5}
\end{equation}
while mass conservation $\nabla\cdot j=0$ reduces to
\begin{equation}
  \frac{\partial\rho}{\partial t} +u^i\partial_i\rho
  +\rho\left(\partial_i+{}_{{}_0}\Gamma^j_{ji}\right)u^i=0,
  \label{e6}
\end{equation}
in the non-relativistic and weak gravity limits. Since Einstein's
field equations can be expressed as
\begin{equation}
  R_{\mu\nu} =- \frac{8\pi G\rho}{c^2}\left(\frac{u_\mu
    u_\nu}{c^2}-\frac{1}{2} g_{\mu\nu}\right) +
  \frac{\Lambda_{DE}G}{c^2}g_{\mu\nu},
  \label{e7}
\end{equation}
$R=4\Lambda_{DE}G/c^2+8\pi G\rho/c^2$, and thus
$\mathfrak{R}\left[c^2R/\Lambda_{DE}G\right]= 
\mathfrak{R}\left[4+8\pi\rho/\Lambda_{DE}\right]$. As is well known,
the only nonvanishing contribution to $R_{\mu\nu}$ in this limit is $R_{00}$,
and Eq.~$(\ref{e7})$ reduces to 
\begin{equation}
  \nabla_i\nabla^i\Phi = 4\pi G\rho + \Lambda_{DE}G.
  \label{e8}
\end{equation}
The last term is small, however, in comparison to $4\pi\rho$, and we
set it to zero from now on. 

\subsection{Spherical geometry and the \textbf{\textit{Evol}}}

We now focus on spherically symmetric
fluid distributions where the fluid rotates about a single rotational
axis. Then using the spherical coordinates
$(r, \theta, \phi)$ where the zenith direction lies along the
rotational axis, the velocity along the polar direction,
$u^\theta(t,r)=0$, vanishes while the density $\rho(t,r)$, the radial
velocity $u^r(t,r)$, and the rotational
(azimuthal) velocity $v(t,r)\equiv u^\phi(t,r)$ are 
functions of $t$ and $r$ only. 

For this fluid Eqs.~$(\ref{e5}), (\ref{e6}),$ and $(\ref{e8})$ reduce
to
\begin{eqnarray}
\frac{Du^r}{\partial t}&=&\frac{L^2}{r^3}-\frac{\partial\>\>\>}{\partial
  r}\left(\Phi + c^2\log\mathfrak{R}\right) - u\frac{D\>\>\>}{\partial
  t}\log\mathfrak{R},
\label{e9}\\
\frac{D\left(L\mathfrak{R}\right)}{\partial t\>\>\>}&=&0,
\label{e10}
\\
\frac{D\rho}{\partial t} &=&-
\frac{\rho}{r^2}\frac{\partial\>\>\>}{\partial r}\left(r^2u\right),
\label{e11}
\\
0&=&\frac{1}{r^2}\frac{\partial\>\>\>}{\partial
  r}\left(r^2\frac{\partial\Phi}{\partial r}\right) -4\pi G\rho,  
\label{e12}
\end{eqnarray}
where $L(t,r)=rv(t,r)$ is the angular momentum of the fluid
while 
\begin{equation}
  \frac{D\>\>\>}{\partial t} = \frac{\partial \>\>\>}{\partial t} +
  u\frac{\partial \>\>\>}{\partial r},
\end{equation}
is the convective derivative. Since $\mathfrak{R}$ depends on $t$ only
implicitly through $\rho$,
\begin{equation}
  \frac{D\mathfrak{R}}{\partial t}=\frac{d\mathfrak{R}}{d\rho}
    \frac{D\rho}{\partial t}.
\end{equation}
But from Eq.~$(\ref{e11})$ we see that $D\rho/\partial t$ is of order
$u$, and the last term in Eq.~$(\ref{e9})$ is of order $u^2$, and thus will
not contribute to our analysis. Similarly, at the length 
scales that we are dealing with and with our interest being on the
structure of galaxies, using the values of $r_H^{*}, v_H^{*},
\alpha_{{}_\Lambda}$, and $\Lambda_{DE}$ given in the introduction, we
find that numerically $\mathfrak{R}\sim 1+\mathcal{O}(10^{-5})$, and
we can set $\mathfrak{R}=1$ in Eq.~$(\ref{e10})$. Angular momentum
conservation follows from Eq.~$(\ref{e10})$, while
Eq.~$(\ref{e11})$ gives mass conservation.

Equations $(\ref{e9})-(\ref{e12})$ give the set of evolution equations
\textbf{\textit{Evol}} for the fluid\footnote{Specifically,
  \textbf{\textit{Evol}} 
  consists of the three evolution equations Eqs. $(\ref{e9})-(\ref{e11})$
  and one constraint equation Eq.~$(\ref{e12})$.} with the solution to
\textbf{\textit{Evol}} denoted by
\begin{equation}
  \mathbf{G}(t,r)=\left(u^r(t,r), L(t,r), \rho(t,r), \Phi(t,r)\right).
\end{equation}
Three out of the four equations that make up
\textbf{\textit{Evol}} are nonlinear, and as such determining a
sufficient set of general boundary conditions needed to obtain a
$\mathbf{G}(t,r)$ is nontrivial. Indeed, this nonlinearity will limit
any definitive comments we can make about the existence of
$\mathbf{G}(t,r)$, or the properties of it. For much of this 
paper we will be guided instead by physical principles. In particular, we
expect on physical grounds that a set of initial conditions
\begin{equation}
  \mathbf{G}_0(r)\equiv\mathbf{G}(0,r)=\left(u^r(0,r), L(0,r),
  \rho(0,r), \Phi(0,r)\right), 
\end{equation}
with $\Phi(0,r)$ given as the solution of Eq.~$(\ref{e12})$, is
needed; \textbf{\textit{Evol}} can then be considered to be the mapping
$\hbox{\textbf{\textit{Evol}}}: \mathbf{G}_0(r) \to 
\mathbf{G}(t,r)$. We also require on physical grounds that as
$r\to\infty$, the three quantities,
$\rho(t,r)\to0$, $u^r(t,r)\to 0$, and $v(t,r)\to 0$, must separately vanish. 

One purpose of this paper is to determine whether the
formation of galaxies with RVCs that agree with observations is
possible under the extended GEOM. As the structure of observed 
galaxies is essentially stationary, one
approach to addressing this question would be to choose a
$\mathbf{G}_0(r)$, solve \textbf{\textit{Evol}} to obtain 
$\mathbf{G}(t,r)$, and then see whether this $\mathbf{G}(t,r)$ evolves in the
$t\to\infty$ limit to
\begin{equation}
  \mathbf{G}_{\infty}(r) = \lim_{t\to\infty}\mathbf{G}(t,r),
  \label{e13}
\end{equation}
a nontrivial, stationary solution of \textbf{\textit{Evol}}. (It
should be noted that not all choices of $\mathbf{G}_0(r)$ need evolve to a
stationary solution of \textbf{\textit{Evol}}, and the
limit in Eq.~$(\ref{e13})$ need not exist. Note also
that since $\mathbf{G}(t,r)$ is a solution to \textbf{\textbf{Evol}}
at each $t>0$, if this limit exists then $\mathbf{G}_{\infty}(r)$
is a stationary solution of \textbf{\textbf{Evol}}.) 
This $\mathbf{G}_{\infty}(t)$ would then be the galaxy predicted to
form under the extended GEOM for this choice of $\mathbf{G}_0(r)$, and
its RVC could be compared to observations. However, while
straightforward, there are a number of issues with this approach. 

\textbf{\textit{Evol}} gives the evolution of \textit{any} initial
distribution of mass and velocities in the spherical geometry. That a
specific choice of $\mathbf{G}_0(r)$ may not result in a stationary
solution of \textbf{\textit{Evol}}, or if it does, may not predict
a galaxy whose RVC agrees with observation, does not mean that the formation
of galaxies with the observed RVCs is not possible under
the extended GEOM. It may simply be that the wrong $\mathbf{G}_0(r)$
was chosen. On the other hand, knowing which $\mathbf{G}_0(r)$ should
be chosen instead is a daunting task. Indeed, given the extreme
nonlinearity of \textbf{\textit{Evol}}, determining the set of
$\mathbf{G}_0(r)$ for which galaxies may be formed under the extended
GEOM, or proving that such a set is empty (as would be expected if
galaxy formation was not possible), is a difficult task. We have
instead taken a different approach, one that focuses on the
stationary solutions of \textbf{\textit{Evol}}.

If a $\mathbf{G}_0(r)$ can be chosen that results in a
$\mathbf{G}_{\infty}(r)$ with a RVC which is consistent with
observations, then such a solution must be a stationary solution of
\textbf{\textit{Evol}}. To determine whether galaxies can form
under the extended GEOM with a RVC that agrees with observation, we 
focus on these stationary solutions. As we show in the next section, stationary
solutions of \textbf{\textit{Evol}} are given by the solution of a
nonlinear, ordinary differential equation, and do not explicitly depend on the
choice of $\mathbf{G}_0(r)$; evolving this $\mathbf{G}_0(r)$ under the
nonlinear evolution equations given by \textbf{\textit{Evol}} can be
avoided. While a stationary solution to \textbf{\textit{Evol}} 
need not be stable, a stability analysis of this solution can be
attained by analyzing the evolution of time-dependent
perturbations about it. Such perturbations naturally linearize
\textbf{\textit{Evol}}, and their evolution is given by linear partial
differential equations whose solutions are tractable. This approach of
finding stationary solutions of \textbf{\textit{Evol}}, and then
analyzing the stability of these solutions is the one we have taken in
this paper.  

\section{The Stationary Limit of \textbf{\textit{Evol}} and Its
  Perturbation} 
\label{Stat}

We turn our attention to the stationary limit of
\textbf{\textit{Evol}}, and the perturbations about it. We
begin by denoting the components of the stationary solution by
$\mathbf{G}_{\infty}(r)=\left(0, L_{\infty}(r),
\rho_{\infty}(r), \Phi_{\infty}(r)\right)$. 
As we are interested in galaxies for which the trajectories of stars
are nearly circular, we have taken $u_{\infty}(r)=0$\footnote{This
  requirement is not too onerous. The stationary radial velocity
  $u_{\infty}(r)=0$ unless $L_{\infty}$ is a constant, which is
  not the case we consider here.}. Moreover, we are in the region
where $2\pi \rho_\infty/\Lambda_{DE} \gg 1$, and we may further approximate
Eq.~$(\ref{e-D})$ as
\begin{equation}
  \mathfrak{D}_{\infty}\left(\frac{8\pi\rho_{\infty}}{\Lambda_{DE}}\right)
  \approx \chi
  \int^\infty_{\frac{8\pi\rho_{\infty}}{\Lambda_{DE}}}s^{-(1+\alpha_{{}_\Lambda})}ds = 
  \frac{\chi}{\alpha_{{}_\Lambda}}\left(\frac{\Lambda_{DE}}{8\pi\rho_{\infty}}\right)^{\alpha_{{}_\Lambda}}. 
  \label{e18}
\end{equation}
It follows that
$\mathfrak{D}_{\infty}\left(8\pi\rho_{\infty}/\Lambda_{DE}\right)\ll 1$,
and thus from Eq. $(\ref{e-R})$,
\begin{equation}
  \mathfrak{R}\left(4+8\pi\rho/\Lambda_{DE}\right)\approx
  1+\frac{1}{2}\mathfrak{D}\left(8\pi\rho/\Lambda_{DE}\right).
\end{equation}

To obtain both the stationary limit of \textbf{\textit{Evol}} and
perturbations about this limit, we perturb the general solution
$\textbf{G}(t,r)$ of \textbf{\textit{Evol}} about
$\mathbf{G}_{\infty}(r)$ by taking $\mathbf{G}(t,r) =
\mathbf{G}_{\infty}(r) + \mathbf{G}_1(t,r)$   
with $\mathbf{G}_1(t,r)\equiv \left(u_1^r(t,r), L_1(t,r), \rho_1(t,r),
\Phi_1(t,r)\right)$ being the perturbation. Keeping to first
order in this perturbation and separating the time-independent
terms from the time dependent ones, we obtain from
\textbf{\textit{Evol}} the equations that determine both
$\mathbf{G}_{\infty}(r)$ and $\mathbf{G}_1(t,r)$. We 
begin with $\mathbf{G}_{\infty}(r)$. 

\subsection{\textbf{\textit{Evol}} in the Stationary Limit}

For $\mathbf{G}_{\infty}(r)$, \textbf{\textit{Evol}} in the nonrelativistic
limit reduces to 
\begin{eqnarray}
  0&=&\frac{L_{\infty}^2}{r^3} - \frac{d\>\>\>}{d
    r}\left(\Phi_{\infty}+\frac{1}{2}c^2\mathfrak{D}_{\infty}\right),
  \label{e14}
  \\
  0&=& \frac{1}{r^2}\frac{d\>\>\>}{dr}
  \left(r^2\frac{d\Phi_{\infty}}{d r}\right)-4\pi 
  G\rho_{\infty},
  \label{e15}
\end{eqnarray}
where $\mathfrak{D}_{\infty}(r)\equiv\mathfrak{D}\left(8\pi
  \rho_{\infty}(r)/\lambda_{DE}\right)$. These two equations can be
combined into one second-order differential equation by multiplying
Eq.~$(\ref{e14})$ by $r^2$ and taking the derivative with respect to
$r$. Equation $(\ref{e15})$ is then used to obtain
\begin{equation}
  \frac{1}{r^2}\frac{d\>\>\>}{dr}\left(\frac{L_{\infty}^2}{r}\right)=4\pi
  G\rho +
  \frac{c^2}{2}\frac{1}{r^2}\frac{d\>\>\>}{dr}
  \left(r^2\frac{d\mathfrak{D}_{\infty}}{dr}\right),
  \label{e16}
\end{equation}
in agreement with \cite{ADS2010b}. 

Treating Eq.~$(\ref{e16})$ as a differential equation for
$\rho_{\infty}(r)$ with a \textit{given} source term
$L_{\infty}(r)$, we find that the solution to Eq.~$(\ref{e16})$
minimizes the time-independent action
\begin{eqnarray}
  S_{\infty} \equiv -\frac{c^2}{16\pi G} 
  \int_0^{r_{II}}&\bigg\{&\frac{c^2}{4}
  \left(\frac{d\mathfrak{D}_{\infty}}{dr}\right)^2
  \nonumber  \\
  &+&\left[\frac{1}{r^2}\frac{d\>\>\>}{dr}\left(\frac{L^2_{\infty}}{r}\right)+
    \frac{\alpha_{{}_\Lambda}}{1-\alpha_{\Lambda}}4\pi
    G\rho_{\infty}\right]\mathfrak{D}_{\infty}\bigg\}
  4\pi r^2dr,
  \label{e17}
\end{eqnarray}
where we have made use of Eq.~$(\ref{e18})$. The integral is to
$r_{II}$ since $2r_{II}$ is the maximum separation between galaxies. 

\subsection{\textbf{\textit{Evol}} for $\mathbf{G}_1(t,r)$}
\label{Perturb}

For $\mathbf{G}_1(t,r)$, \textbf{\textit{Evol}} reduces to
\begin{eqnarray}
  \frac{\partial u_1^r}{\partial t} &=& \frac{2}{r^3} L_{\infty}L_1 -
  \frac{\partial\>\>\>}{\partial r}\left[\Phi_1 -
  \frac{c^2}{2}\chi \left(\frac{\Lambda_{DE}}{8\pi\rho_{\infty}}\right)^{\alpha_{\Lambda}+1} \frac{8\pi\rho_1}{\Lambda_{DE}}\right],
  \label{e19}
  \\
  \frac{\partial L_1}{\partial t} &=&-u_1^r \frac{d
    L_{\infty}}{d r},
  \label{e20}
  \\
  \frac{\partial \rho_1}{\partial t} &=&-\frac{1}{r^2}
  \frac{\partial\>\>\>}{\partial r} \left(r^2\rho_{\infty} u_1^r\right),
  \label{e21}
  \\
  0&=& \frac{1}{r^2}\frac{d\>\>\>}{dr}
  \left(r^2\frac{\partial\Phi_1}{\partial r}\right)-4\pi 
  G\rho_1,
  \label{e22}
\end{eqnarray}
in the nonrelativistic limit. As with the stationary equations, these
four equations can be reduced to a single second-order differential
equation. 

Taking the derivative  of Eq.~$(\ref{e12})$ with respect to $t$ and
making use of Eq.~$(\ref{e11})$, we find that
\begin{equation}
  \frac{G\dot{M}(t)}{r^2} = \frac{\partial^2 \Phi}{\partial t\partial r} +
  4\pi G\rho(t,r) u^r(t,r),
  \label{e23}
\end{equation}
where $\dot{M}(t)$ is an arbitrary function of time only. Next, taking the
derivative of Eq.~$(\ref{e19})$ with respect to $t$, and making use of
Eqs.~$(\ref{e20})$, $(\ref{e21})$ and $(\ref{e23})$, we arrive at
\begin{eqnarray}
-\frac{G\dot{M}(t)}{r^2} = &{}&\frac{\partial^2 u_1^r}{\partial t^2} +
\frac{1}{2}c^2\chi\frac{\partial\>\>\>}{\partial
  r}\left\{
  \left(\frac{\Lambda_{DE}}{8\pi\rho_{\infty}}\right)^{1+\alpha_{{}_\Lambda}} 
  \frac{1}{r^2}
  \frac{\partial\>\>\>}{\partial
    r}\left[r^2\left(\frac{8\pi \rho_{\infty}}{\Lambda_{DE}}\right)u_1^r\right]\right\}
\nonumber
\\
&+&
\left[\frac{1}{r^3}\frac{dL_{\infty}^2}{dr}-4\pi
  G\rho_{\infty}\right]u_1^r,
\label{e24}
\end{eqnarray}
since $\rho(t,r)u^r(t,r)\approx\rho_{\infty}u_1^r(t,r)$ to first order.
The solution of this differential equation for $u_1^r$ also minimizes an action,
\begin{eqnarray}
S_1\equiv \frac{c^2}{16\pi G} 
\int_0^{r_{II}}&\bigg\{&
\frac{4\pi\rho_{\infty}}{\Lambda_{DE}}\left(\frac{\partial
  u_1^r}{\partial t}\right)^2
+
\nonumber
\\
&{}&\frac{c^2\chi}{4}\left(\frac{\Lambda_{DE}}{8\pi\rho_{\infty}}\right)^{\alpha_{{}_\Lambda}+1}
\frac{1}{r^4}\left(\frac{\partial\>\>\>}{\partial
  r}\left[\frac{8\pi\rho_{\infty}}{\Lambda_{DE}}r^2u_1^r\right]\right)^2-
\nonumber
\\
&{}&\frac{4\pi
  \rho_{\infty}}{\Lambda_{DE}}\left(\frac{1}{r^3}\frac{dL^2_{\infty}}{dr}-4\pi
G\rho_{\infty}\right){u_1^r}^2
-\frac{8\pi G\rho_{\infty}}{\Lambda_{DE} r^2}\dot{M}u_1^r
\bigg\} 4\pi r^2dr dt,
\nonumber
\\
\end{eqnarray}
for a given $L_{\infty}(r)$.

As the current density along the radial direction
$j^r(t,r)=\rho(t,r)u^r(t,r)$, the mass flux through a sphere
$Sph(R)$ with radius $R$ about the center of the galaxy is
\begin{equation}
  \int_{Sph(R)} \vec{\jmath}\cdot d\vec{A} = 4\pi R^2\rho(t,R)u^r(t,R).
\end{equation}
From Eq.~$(\ref{e23})$, this flux is
\begin{equation}
  \int_{Sph(R)} \vec{\jmath}\cdot d\vec{A} = \dot{M}(t)
  -\frac{\partial\>\>\>}{\partial t}\left(\frac{R^2}{G}\frac{\partial
    \Phi}{\partial r}\bigg\vert_{r=R}\right).
\end{equation}
When $\dot{M}(t)>0$ there is a flux of mass leaving the center of the
galaxy. and thus the mass of the galaxy would be increasing at the
rate $\dot{M}(t)$ even in the limit $R\to0$. On the other hand, when
$\dot{M}(t)<0$ there is a flux of mass entering the center of the
galaxy, and thus the mass of the galaxy would be decreasing at the
rate of $\dot{M}(t)$. There will thus be an essential singularity at
the center of the galaxy that would either inject mass into the
galaxy, or remove mass from it. In either case, the total mass of the
galaxy would not be conserved if $\dot{M}(t)\ne0$. As
Eq.~$(\ref{e11})$ asserts that mass is in fact conserved, we set
$\dot{M}(t)=0$.  

\subsection{The Role of Angular Momentum}

In the passage from $\mathbf{G}(t,r)$ to $\mathbf{G}_{\infty}(r)$,
\textbf{\textit{Evol}} reduces from four equations determining four
functions to two equations determining three functions;
\textbf{\textit{Evol}} thus
becomes a system of \textit{underdetermined}  
differential equations in the stationary limit. This can be seen
explicitly in Eq.~$(\ref{e16})$ where 
$\rho_{\infty}(r)$ is only determined once $L_{\infty}(r)$ is
known. This underdeterminacy is expected, and is consistent with
observations.  

Suppose instead that \textbf{\textit{Evol}} reduces in the stationary
limit to a system of three
differential equations for the three non-zero components of
$\mathbf{G}_{\infty}(r)$. This system of differential equations
would be complete, and could then be solved without reference
to the initial conditions $\mathbf{G}_0(r)$; they would
only have to satisfy the same the boundary conditions at $r=0$ and $r\to
\infty$ that are required of $\mathbf{G}(t,r)$. The
solutions of these differential equations will include three arbitrary
constants that would then be determined by these boundary
conditions. It would then follow that the stationary galaxies predicted
by the  extended GEOM would all be the same irrespective of the choice of
initial condition $\mathbf{G}_{0}(r)$. This certainly is not what is observed. 

That the set of differential equations given by
\textbf{\textit{Evol}} is underdetermined means that
$\mathbf{G}_{\infty}(r)$ depends indirectly on the choice of initial
conditions $\mathbf{G}_0(r)$. A choice of $\mathbf{G}_0(r)$ will,
after evolving with \textbf{\textit{Evol}}, give a $L_{\infty}(r)$
that will, through the solution of Eq.~$(\ref{e16})$, give
$\rho_{\infty}(r)$ as well. Indeed, the dependence of
Eq.~$(\ref{e16})$ on $L_{\infty}(r)$, and its role as the driving term for determining
$\rho_{\infty}(r)$ underscores the important role that angular
momentum plays in the formation of galaxies. However, since the evolution
of $\mathbf{G}_0(r)$ to get a $L_{\infty}(r)$ would require the solution of
\textbf{\textit{Evol}}, and since such a solution would also give
$\rho_{\infty}(r)$ in the first place, one can question the
usefulness of focusing on the stationary solutions of
\textbf{\textit{Evol}} and Eq.~$(\ref{e16})$. In the end, it comes
down to the choice of angular momentum for the fluid, and when 
this choice is made. 

We can certainly make this choice at $t=0$ by choosing a specific
$\mathbf{G}_0(r)$, and this choice may then result in a 
$L_{\infty}(r)$ in the stationary limit after evolution under
\textbf{\textit{Evol}}. We can also make this 
choice at the stationary limit by choosing a $L_{\infty}(r)$
directly, \textit{with the expectation} that, since the angular momentum of the
system is conserved, there exists some choice of
$\mathbf{G}_0(r)$ that will give this $L_{\infty}(r)$  after
evolving with \textbf{\textit{Evol}}. By making the choice at the
stationary limit we circumvent the difficulty of solving the nonlinear
partial differential equations in \textbf{\textit{Evol}}. Moreover, with an
appropriate choice of $L_{\infty}(r)$ we will also be able to
determine whether it is possible for the extended GEOM to form
galaxies with RVCs that agree with observation. This choice of
$L_{{}_\infty}$ can be determined using $S_{{}_\infty}$ and the fact
for given a stationary $L_{{}_\infty}$ the stationary density
$\rho_{{}_\infty}$ must minimize $S_{{}_\infty}$.

The $v_{\infty}(r)$ considered here in Eq.~$(\ref{e1})$ is a natural
generalization of $v^{\hbox{\scriptsize{ideal}}}(r)$. It has a
well-defined asymptotic behavior in both the $x\to 0$ and $x\to
\infty$ limits, and the two behaviors are smoothly joined together. It depends
on four parameters, $r_H^{*}, v_H^{*}, q, p$, and as $q$ and $p$
give the power law behavior of $v_{\infty}(r)$ in the asymptotic
limits $x\to0$ and $x\to\infty$ respectively, each parameter has a definite physical
interpretation. Two of the parameters, $r_H^{*}$ and $v_H^{*}$, are
set by observations, and are considered fixed. The remaining two 
parameters, $q$ and $p$, are considered to be free, and forms a
two-dimensional parameter space $\mathcal{P}$. As each chosen $q$ and
$p$ will give a different RVC, this $v_{\infty}(r)$ describes a
class of velocity profiles, each similar in form, and, since by
construction $v_{\infty}(x=1)=v_H^{*}$, all of whom can be compared with one
another. We will, in addition, require that 
$v_{\infty}(x)\to 0$ as $x\to 0$ and $x\to \infty$; this in turn requires
that $q>0$ and $p> 0$. Newtonian gravity, on the other hand, 
would set $q=1$ and $p=1/2$; we follow suit and limit $q\le1$ and
$p\le 1/2$. The parameter space is thus restricted to the strip
$\mathcal{P}=\{(q,p): 0<q\le 1, 0< p\le1/2\}$. The main focus of this
paper is determining the region of $\mathcal{P}$ for which galaxies
with velocity profiles $v_{\infty}(x)$ can be formed under the
extended GEOM. 

We choose $L_{\infty}(r) = rv_{\infty}(r)$. Then each $q$ and
$p$ will give a specific $L_{\infty}(r;q,p)$, and 
through the solution of Eq.~$(\ref{e16})$, a density profile
$\rho_{\infty}(r;q,p)$ for a galaxy in the stationary limit. But while
each choice of $q$ and $p$ may ultimately result in a
$\rho_{\infty}$, such a $\rho_{\infty}$ need not minimize
$S_{\infty}$. By evaluating $S_{\infty}$ at
$\rho_{\infty}(r;q,p)$ and $L_{\infty}(r;q,p)$, and minimizing the action with
respect to the parameters, we are able to determine the values of $q$
and $p$ that do produce a $L_{\infty}(r;q,p)$ and a
$\rho_{\infty}(r;q,p)$ which minimizes the action. It is for these
values of $q$ and $p$ that a stationary galaxy with a RVC given by
$v_{\infty}(x)$ can be formed under the extended GEOM. Importantly,
if no such $q$ and $p$ can be found, then such galaxies could not form
under the extended GEOM.   

\section{Stationary Solutions}
\label{Solu}

We now turn our attention to finding solutions to
Eq.~$(\ref{e16})$. This is possible due to the two drastically
different length scales in the theory, $r_H^{*}$ and
$\lambda_{DE}$. A straightforward use of them results in a small
parameter that allows for a perturbative solution of
Eq.~$(\ref{e16})$. This parameter multiplies the 
highest-order derivative of the differential equation, however, and
thus an application of perturbation theory results in a
reduction of the order of this differential equation. The perturbation
theory is therefore singular, and thus techniques from boundary layer
theory (see Ch.~9 of \cite{Bend1978}) will have to be
applied. However, there are such significant differences between
Eq.~$(\ref{e25})$ and its boundary conditions, and the differential
equations and boundary conditions analyzed in \cite{Bend1978} that 
the analysis outlined and the terminology used in \cite{Bend1978}
cannot be directly applied here. Rather, they will instead
serve as guidance for our analysis. Indeed, like the boundary layer
analysis \cite{Bend1978}, perturbative solutions for our differential
equation must be found in two or more regions of space, and a
consistent solution only exists if these regions
overlap. In our case, we will find both the leading and the
first-order perturbation solutions of Eq.~$(\ref{e25})$ within the
galactic hub and within the galactic disk, and then we will determine for
which $v_{\infty}(x)$ these two regions overlap. Unlike the
differential equations considered in \cite{Bend1978}, however, one of
the boundary conditions for Eq.~$(\ref{e25})$ is in the limit
$x\to\infty$, and thus use of the outer- and inner- terminology
used in \cite{Bend1978} would be confusing at best, and we do not use
it here. More importantly, with $r_H^{*}$ and $\lambda_{DE}$ we have two
different length scores, and this will allow for a different approach 
to finding the inner-limit solution than that described in
\cite{Bend1978}. We begin at the $r_H^{*}$ length scale.

\subsection{The Solution in the Galactic Hub Region}

In this region we make use of the length scale $r_H^{*}$, and take
$x=r/r_H^{*}$. Then defining $\widehat{v}_{\infty}(x)\equiv
v_{\infty}(x)/v_H^{*}$ and  
$\Upsilon(x)=(\rho_{\infty}/\rho_H^{*})^{-\alpha_{{}_\Lambda}}$ with 
\begin{equation}
  \rho_H^{*} = \frac{3{v_H^{*}}^2}{4\pi G {r_H^{*}}^2},
\end{equation}
Eq.~$(\ref{e16})$ becomes
\begin{equation}
    F(x)=\Upsilon^{-1/\alpha_{{}_\Lambda}}
    +\frac{\epsilon^2}{x^2}\frac{d\>\>\>}{dx}\left(x^2\frac{d\Upsilon}{dx}\right), 
    \label{e25}
\end{equation}
where
\begin{equation}
  F(x) = \frac{1}{3x^2}\frac{d\>\>\>}{dx}\left(x\widehat{v}_{\infty}^2(x)\right),
\end{equation}
while
\begin{equation}
  \epsilon^2 =
  \frac{\chi}{\alpha}\left(\frac{c^2}{6{v_H^{*}}^2}\right)^{1+\alpha_{{}_\Lambda}}
  \left(\frac{{r_H^{*}}^2}{\lambda_{DE}^2}\right)^{\alpha_{{}_\Lambda}}.
\end{equation}
The values for $r_H^{*}, v_H^{*}$ and $\lambda_{DE}$ given in
the introduction are used to obtain $\epsilon=6.131\times 10^{-3}$
after evaluating $\chi$ at $\alpha_{{}_\Lambda}=1.56_{\pm 0.10}$. Using
$\epsilon$ as an expansion parameter, we first take the outer limit
$\epsilon\to0$ \cite{Bend1978}, and expand $\Upsilon(x)$ to first
order in $\epsilon^2$: $\Upsilon(x) = 
\Upsilon_0(x)+\epsilon^2\Upsilon_1(x)$. Equation $(\ref{e25})$ gives
for the leading term 
\begin{equation}
  \Upsilon_0(x)=\left[F(x)\right]^{-\alpha_{{}_\Lambda}},
\end{equation}
and like the nonlinear Carrier equation \cite{Bend1978} the reduction of
order in Eq.~$(\ref{e25})$ results in an algebraic equation that is
  easily solved. Indeed, the resultant equation in this limit is
  algebraic to all orders in the perturbation expansion. In
  particular, it gives for the first-order perturbation 
\begin{equation}
  \Upsilon_1(x)=
  \frac{\alpha}{\left[F(x)\right]^{\alpha_{{}_\Lambda}+1}}\frac{1}{x^2}\frac{d\>\>\>}{dx}
  \left(x^2\frac{d\>\>\>}{dx}\left[F(x)\right]^{-\alpha_{{}_\Lambda}}\right).
  \label{e26}
\end{equation}
The resulting density is
\begin{equation}
  \rho_{\infty}^{\hbox{\scriptsize{hub}}}(x) =
  \frac{F(x)\rho_H^{*}}{\left[1+\epsilon^2\Upsilon_1(x)/\Upsilon_0(x)\right]^{1/\alpha_{{}_\Lambda}}}.
  \label{e27}
\end{equation}

Equation $(\ref{e27})$ is valid for values of $x$ for which 
$\epsilon^2\vert\Upsilon_1(x)\vert<\vert\Upsilon_0(x)\vert$, or
equivalently, when $E_{\hbox{\scriptsize{hub}}}(x)\equiv\epsilon^2\vert
\Upsilon_1(x)/\Upsilon_0(x)\vert < 1$. This condition establishes the
region $\mathcal{R}_{\hbox{\scriptsize{hub}}}$ of space for which
Eq.~$(\ref{e27})$ is valid. We will see below that
$\mathcal{R}_{\hbox{\scriptsize{hub}}}$ encompasses the region
around $r=0$, and thus corresponds to the galactic hub.  Indeed, this
can be seen directly by setting $\epsilon=0$ in Eq.~$(\ref{e27})$;
$\rho_{\infty}(r)$ then reduces to the Newtonian result. 

\subsection{The Solution in the Galactic Disk Region}
\label{Inner}

In this region we make use of the length scale $\lambda_{DE}$, and
take $\bar{x}\equiv r/(\chi^{1/2}\lambda_{DE})$. Then defining
$\bar{y}(\bar{x}) \equiv 8\pi\rho_{\infty}(\bar{x})/\Lambda_{DE}$, and 
\begin{equation}
  \bar{F}(\bar{x})\equiv
  \frac{2}{\chi}\left(\frac{{v_H^{*}}^2}{c^2}\right)\frac{1}{\bar{x}^2}
  \frac{d\>\>\>}{d\bar{x}}\left(\bar{x}\widehat{v}_{\infty}^2(\bar{x})\right),
\end{equation}
Eq.~$(\ref{e16})$ becomes
\begin{equation}
 \bar{F}(\bar{x})=\bar{y}+\frac{1}{\alpha_{\Lambda}}\frac{1}{\bar{x}^2}
   \frac{d\>\>\>}{d\bar{x}}\left(\bar{x}^2\frac{dy^{-\alpha_{{}_\Lambda}}}{d\bar{x}\>\>\>}\right).
\label{e28}
\end{equation}
For $\bar{x}_H\equiv r_H^{*}/\chi^{1/2}\lambda_{DE}=7.392\times
10^{-7}$, 
\begin{equation}
  \widehat{v}_{\infty}(\bar{x}) =
  \sqrt{\frac{1+p/q}{\bar{x}^{2(q+p)}+\bar{x}_H^{2(q+p)}p/q}}\bar{x}^q\bar{x}_H^p,
\end{equation}
after using $x=\bar{x}/\bar{x}_H$, We see that when
$\bar{x}\sim 1$, $\widehat{v}_{\infty}\sim \bar{x}_H^p$, and $\bar{F}\sim
\bar{x}_H^{2p}{v_H^{*}}^2/c^2$. This leads us to take  
\begin{equation}
  \bar{y}(\bar{x}) =
  \bar{y}_a(\bar{x})+\bar{x}_H^{2p}\frac{{v_H^{*}}^2}{c^2} \bar{y}_1(\bar{x}),
\end{equation}
and expand Eq.~$(\ref{e16})$ to first order in $\bar{x}_H^{2p}{v_H^{*}}^2/c^2$. The
leading term $\bar{y}_a$ is given by 
\begin{equation}
  0=\bar{y}_a+ \frac{1}{\alpha_{{}_\Lambda}}\frac{1}{\bar{x}^2}\frac{d\>\>\>}{d\bar{x}}
  \left(\bar{x}^2\frac{d\bar{y}_a^{-\alpha_{{}_\Lambda}}}{d\bar{x}\quad}\right).
  \label{e29}
\end{equation}
The solution to Eq.~$(\ref{e29})$ was found in \cite{ADS2010b};
the details of how this was done can be found there. Here, we will only
need the following
\begin{equation}
  \bar{y}_a(\bar{x})=\frac{\Sigma}{\bar{x}^{\frac{2}{\alpha_{{}_\Lambda}+1}}},
  \label{e30}
\end{equation}
where
\begin{equation}
  \Sigma^{\alpha_{{}_\Lambda}+1}=-\frac{2(1+3\alpha_{{}_\Lambda})}{(1+\alpha_{{}_\Lambda})^2},
  \qquad \hbox{while}\qquad \Sigma =
  \left[\frac{2(1+3\alpha_{{}_\Lambda})}{(1+\alpha_{{}_\Lambda})^2}\right]^{\frac{1}{1+\alpha_{{}_\Lambda}}}.
  \label{e30b}
\end{equation}
For the first-order perturbation $\bar{y}_1$, on the other hand,
Eq.~$(\ref{e16})$ gives 
\begin{equation}
\bar{x}_H^{-2p}\frac{c^2}{{v_H^{*}}^2}\bar{F}(\bar{x})=\bar{y}_1+
\frac{1}{\vert\Sigma^{(\alpha_{{}_\Lambda}+1)}\vert}\frac{1}{\bar{x}^2}\frac{d\>\>\>}{d\bar{x}} 
  \left[\bar{x}^2\frac{d\>\>\>}{d\bar{x}}\left(\bar{x}^2\bar{y}_1\right)\right].
  \label{e31}
\end{equation}
The extent of the region $\mathcal{R}_{\hbox{\scriptsize{disk}}}$ where
this perturbative expansion is valid is determined by the condition 
$E_{\hbox{\scriptsize{disk}}}(\bar{x})\equiv \bar{x}_H^{2p}(v_H^{*})/c)^2 \vert
\bar{y}_1(\bar{x})/\bar{y}_a(\bar{x})\vert<1$. As we will see
below, this region excludes the point $r=0$, and thus
corresponds to the galactic disk.

Equation $(\ref{e31})$ is straightforwardly solved to give
\begin{eqnarray}
  \bar{y}(\bar{x}) = &{}&\frac{\Sigma}{\bar{x}^{\frac{2}{\alpha_{{}_\Lambda}+1}}} +
  \bar{x}_H^{2p}\frac{{v_H^{*}}^2}{c^2}
  \left[\bar{C}_{cos}j_c(\bar{x}/\bar{x}_0)+
    \bar{C}_{sin}j_s(\bar{x}/\bar{x}_0)\right]-
  \nonumber
  \\
  &{}&\frac{2}{\nu\chi}\frac{{v_H^{*}}^2}{c^2}
  \frac{\vert\Sigma^{1+\alpha_{{}_\Lambda}}\vert}{\bar{x}^{5/2}}
  \int_{\bar{x}_0}^{\bar{x}}\frac{d\>\>\>}{d\bar{s}}\left(\bar{s}\widehat{v}_{\infty}^2\right)
  \sin(\nu\log{\bar{s}/\bar{x}})\frac{d\bar{s}}{\sqrt{\bar{s}}},
\end{eqnarray}
where $\nu = \sqrt{\vert\Sigma^{\alpha_{{}_\Lambda}+1}\vert-1/4}$,
\begin{equation}
  j_c(x) = \frac{1}{x^{5/2}}\cos(\nu\log x), \hbox{ and }
  j_s(x) = \frac{1}{x^{5/2}}\sin(\nu\log x).
\end{equation}
Here $\bar{x}_0$ is a point in the intersection of the regions
$\mathcal{R}_{\hbox{\scriptsize{hub}}}$ and
$\mathcal{R}_{\hbox{\scriptsize{disk}}}$. In the next section we will
determine the region of $\mathcal{P}$ where this intersection is
nonempty. For now, we will assume that we are working in this
region. We then make use of the following expansion to evaluate the
integral    
\begin{equation}
  \widehat{v}_{\infty}^2(\bar{x}) =
  (1+p/q)\left(\frac{\bar{x}_H}{\bar{x}_c}\right)^{2p}\sum_{n=0}^\infty
  (-1)^n
  \left[\left[\frac{\bar{x}}{\bar{x}_c}\right]^{Z_{qp}^n}\theta(\bar{x}_c-\bar{x})+
    \left[\frac{\bar{x}_c}{\bar{x}}\right]^{Z_{pq}^n}\theta(\bar{x}-\bar{x}_c)\right],
  \label{e31b}
\end{equation}
where $\bar{x}^{2(q+p)}_c=(p/q)\bar{x}_H^{2(q+p)}$,
$Z_{qp}^n=2[n(q+1)+np]$, $Z_{pq}^n=2[n(p+1)+nq]$, and $\theta(x)$ is
the Heaviside function. The resulting density is 
\begin{eqnarray}
  \rho_{\infty}^{\hbox{\scriptsize{disk}}}(x)
  =\rho_H^{*}\Sigma\left(\frac{\alpha\epsilon^2}{x^2}\right)^{\frac{1}{\alpha_{{}_\Lambda}+1}}+
  C_{cos}j_c\left(\frac{x}{x_0}\right)+C_{sin}j_s\left(\frac{x}{x_0}\right)+
  \rho_p^{\hbox{\scriptsize{disk}}}(x),
  \label{e32}
\end{eqnarray}
where the constants
\begin{eqnarray}
  C_{cos} &=& \rho_{\infty}^{\hbox{\scriptsize{hub}}}(x_0^{-})
  -\rho_H^{*}\Sigma\left(\frac{\alpha_{{}_\Lambda}\epsilon^2}{x_0^2}\right)^{\frac{1}{\alpha_{{}_\Lambda}+1}},
  \\
  \nu C_{sin} &=&
  x\frac{d\rho^{\hbox{\scriptsize{hub}}}_{\infty}}{dx\>\>\>}\bigg\vert_{x_0^{-}}+
  \frac{5}{2}\rho_{\infty}^{\hbox{\scriptsize{hub}}}(x_0^{-})
  -\frac{1}{2}\rho_H^{*}\Sigma\left(\frac{1+5\alpha_{{}_\Lambda}}{1+\alpha_{{}_\Lambda}}\right)
  \left(\frac{\alpha_{{}_\Lambda}\epsilon^2}{x_0^2}\right)^{\frac{1}{\alpha_{{}_\Lambda}+1}},
\end{eqnarray} 
are determined by requiring the density to be smooth at the
transition point $x_0=\bar{x}_0/\bar{x}_H$ between the regions
$\mathcal{R}_{\hbox{\scriptsize{hub}}}$ and
$\mathcal{R}_{\hbox{\scriptsize{disk}}}$, 
\begin{equation}
  \rho_{\infty}^{\hbox{\scriptsize{hub}}}(x_0^{-}) =
  \rho_{\infty}^{\hbox{\scriptsize{disk}}}(x_0^{+}), \qquad
  \frac{d\rho_{\infty}^{\hbox{\scriptsize{hub}}}}{dx\>\>\>}\bigg\vert_{x_0^{-}}
  = 
  \frac{d\rho_{\infty}^{\hbox{\scriptsize{disk}}}}{dx\>\>\>}\bigg\vert_{x_0^{+}}.
\end{equation}
As for the particular solution $\rho_{\hbox{\scriptsize{part}}}^{\hbox{\scriptsize{disk}}}(x)$, when $x_c<x_0$, 
\begin{eqnarray}
  \frac{\rho_{\hbox{\scriptsize{part}}}^{\hbox{\scriptsize{disk}}}(x)}{\rho_H^{*}} =&{}&
  \frac{1}{3}\frac{(1+p/q)}{x_c^{2(p+1)}}
  \sum_{n=0}^\infty
  \frac{(-1)^n \vert\Sigma^{\alpha_{{}_\Lambda}+1}\vert(1-Z_{pq}^n)}
       {\vert\Sigma^{\alpha_{{}_\Lambda}+1}\vert+Z_{pq}^n(Z_{pq}^n-1)}
       \bigg\{\left[\frac{x_c}{x}\right]^{Z_{pq}^n+2}-
       \nonumber
       \\
       &{}&\left[j_c\left(\frac{x}{x_0}\right)+ \frac{1}{\nu}\left[\frac{1}{2}-Z_{pq}^n\right]j_s\left(\frac{x}{x_0}\right)\right]\left[\frac{x_c}{x_0}\right]^{Z_{pq}^n+2}
       \bigg\},
       \label{e32b}
\end{eqnarray}
while when $x_0\le x\le x_c$,
\begin{eqnarray}
  \frac{\rho_{\hbox{\scriptsize{part}}}^{\hbox{\scriptsize{disk}}}(x)}{\rho_H^{*}}= &{}&\frac{1}{3}\frac{(1+p/q)}{x_c^{2(p+1)}}
  \sum_{n=0}^\infty\frac{(-1)^n\vert\Sigma^{\alpha_{{}_\Lambda}+1}\vert(1+Z_{qp}^n)}
      {\vert\Sigma^{\alpha_{{}_\Lambda}+1}\vert+Z_{qp}^n(Z_{qp}^n+1)}
      \bigg\{\left[\frac{x}{x_c}\right]^{Z_{qp}^n-2} -
      \nonumber
      \\
      &{}&\left[j_c\left(\frac{x}{x_0}\right)+\frac{1}{\nu}\left[\frac{1}{2}+Z_{qp}^n\right]j_s\left(\frac{x}{x_0}\right)\right]
      \left[\frac{x_0}{x_c}\right]^{Z_{qp}^n-2}
      \bigg\}.
      \label{e32c}
\end{eqnarray}
Finally, when $x_0\le x_c\le x$,
\begin{eqnarray}
\frac{\rho_{\hbox{\scriptsize{part}}}^{\hbox{\scriptsize{disk}}}(x)}{\rho_H^{*}} = &{}&\frac{1}{3}\frac{(1+p/q)}{x_c^{2(p+1)}}
  \sum_{n=0}^\infty(-1)^n \vert\Sigma^{\alpha_{{}_\Lambda}+1}\vert\Bigg\{
  \nonumber
  \\
  &{}&
  \frac{(1-Z_{pq}^n)}{\vert\Sigma^{\alpha_{{}_\Lambda}+1}\vert+Z_{pq}^n(Z_{pq}^n-1)}
  \left[\frac{x_c}{x}\right]^{Z_{pq}^n+2} +
  \nonumber
  \\
  &{}&\Bigg[
         \left[\frac{(1+Z_{qp}^n)}{\vert\Sigma^{\alpha_{{}_\Lambda}+1}\vert+Z_{qp}^n(Z_{qp}^n+1)}
         -
         \frac{(1-Z_{pq}^n)}{\vert\Sigma^{\alpha_{{}_\Lambda}+1}\vert+Z_{pq}^n(Z_{pq}^n-1)}
         \right]j_c\left(\frac{x}{x_c}\right)-
         \nonumber
         \\
        &{}&\frac{1}{\nu}
         \left[\frac{\vert\Sigma^{\alpha_{{}_\Lambda}+1}\vert-\frac{1}{2}\left(1+Z_{qp}^n\right)}
             {\vert\Sigma^{\alpha_{{}_\Lambda}+1}\vert+Z_{qp}^n(Z_{qp}^n+1)}
         -
         \frac{\vert\Sigma^{1+\alpha_{{}_\Lambda}}\vert-\frac{1}{2}(1-Z_{pq}^n)}{\vert\Sigma^{\alpha_{{}_\Lambda}+1}\vert+Z_{pq}^n(Z_{pq}^n-1)}
         \right]j_s\left(\frac{x}{x_c}\right)
         \Bigg] 
  \nonumber
  \\
  &{}&
  -(1+Z_{qp}^n)\frac{\left[j_c\left(\frac{x}{x_0}\right)+\frac{1}{\nu}\left[\frac{1}{2}+Z_{qp}^n\right]j_s\left(\frac{x}{x_0}\right)\right]}{\vert\Sigma^{\alpha_{{}_\Lambda}+1}\vert+Z_{qp}^n(Z_{qp}^n+1)}
  \left[\frac{x_0}{x_c}\right]^{Z_{qp}^n-2}
  \Bigg\}.
  \nonumber
  \\
  \label{e32d}
\end{eqnarray}

\subsection{Consistent Solutions}

The region where $\rho_{\infty}^{\hbox{\scriptsize{hub}}}$ is valid is given by
$\mathcal{R}_{\hbox{\scriptsize{hub}}}=\{x:E_{\hbox{\scriptsize{hub}}}(x)<1\}$;
this region is simply connected, and 
$\mathcal{R}_{\hbox{\scriptsize{hub}}}=(x_L^{\hbox{\scriptsize{hub}}},
x_U^{\hbox{\scriptsize{hub}}})$. Similarly, the region where
$\rho_{\infty}^{\hbox{\scriptsize{disk}}}$ is valid is given by
$\mathcal{R}_{\hbox{\scriptsize{disk}}}
= \{x:E_{\hbox{\scriptsize{disk}}}(x)<1\}$; this region is also simply
connected with
$\mathcal{R}_{\hbox{\scriptsize{disk}}}=(x_L^{\hbox{\scriptsize{disk}}}, 
x_U^{\hbox{\scriptsize{disk}}})$.  A consistent
solution $\rho_{\infty}(x)$ to Eq.~$(\ref{e16})$ exists when
the two regions overlap,
$\mathcal{R}_{\hbox{\scriptsize{hub}}}\cap\mathcal{R}_{\hbox{\scriptsize{disk}}}\ne\emptyset$ 
\cite{Bend1978}. It is only in this case that a $x_0$ can be chosen,
and the arbitrary constants $\bar{C}_{cos}$ and $\bar{C}_{sin}$ in the
homogenous solution to Eq.~$(\ref{e16})$ be
determined. The focus of this subsection is on determining both
$\mathcal{R}_{\hbox{\scriptsize{hub}}}$ and
$\mathcal{R}_{\hbox{\scriptsize{disk}}}$, along with the region
$\mathcal{P}_\cap\subset\mathcal{P}$ for which their intersection
is not empty: $\mathcal{P}_\cap = \{(q,p)\in\mathcal{P} : 
\mathcal{R}_{\hbox{\scriptsize{hub}}}\cap\mathcal{R}_{\hbox{\scriptsize{disk}}}\ne\emptyset\}$.   

Considering first the limit $x\to0$, we find that
\begin{equation}
 E_{\hbox{\scriptsize{hub}}}(x)\approx 2\alpha_{{}_\Lambda}\epsilon^2
    \frac{(1-q)(1+2\alpha_{{}_\Lambda}(1-q))}{\left[(1+q/p)(1+2q)/3\right]^{1+\alpha_{{}_\Lambda}}}
    x^{-2[q(1+\alpha_{{}_\Lambda})-\alpha_{{}_\Lambda}]}.
\end{equation}
 Then $x^{\hbox{\scriptsize{hub}}}_L=0$ for
 $q\le\alpha_{{}_\Lambda}/(\alpha_{{}_\Lambda}+1)$ and $q=1$, while when
 $\alpha_{{}_\Lambda}/(\alpha_{{}_\Lambda}+1)<q<1$,
 \begin{equation}
   x_L^{\hbox{\scriptsize{hub}}}=\left(2\alpha_{{}_\Lambda}\epsilon^2
     \frac{(1-q)(1+2\alpha_{{}_\Lambda}(1-q))}{\left[(1+q/p)(1+2q)/3\right]^{1+\alpha_{{}_\Lambda}}}\right)^{\frac{1}{2[q(1+\alpha_{{}_\Lambda})-\alpha_{{}_\Lambda}]}}.
 \end{equation}
The largest that $x_L^{\hbox{\scriptsize{hub}}}$ can be
is $2.69\times 10^{-4}$ or roughly 3.18 pc. In the
$x\to\infty$ limit, on the other
hand, $\rho_{\infty}^{\hbox{\scriptsize{disk}}}(x)\sim x^{-2(p+1)}$
while $\rho_a(x)\sim x^{-2/(\alpha_{\infty}+1)}$. Then 
$E_{\hbox{\scriptsize{disk}}}(x)\sim
x^{-2(p+\alpha_{{}_\Lambda}/(\alpha_{{}_\Lambda}+1))}\to 0$ as 
$x\to\infty$ for all $p$, and thus
$\mathcal{R}_{\hbox{\scriptsize{disk}}}=(x_L^{\hbox{\scriptsize{disk}}},
\infty)$. 

Both $x_U^{\hbox{\scriptsize{hub}}}$ and $x_L^{\hbox{\scriptsize{disk}}}$
are found numerically using the following process. For $x_U^{\hbox{\scriptsize{hub}}}$, a value of
$x_{\hbox{\scriptsize{trial}}}$ is chosen, and
$E_{\hbox{\scriptsize{hub}}}(x_{\hbox{\scriptsize{trial}}})$ is 
calculated. If $E_{\hbox{\scriptsize{hub}}}(x_{\hbox{\scriptsize{trial}}})>1$, the
value of $x_{\hbox{\scriptsize{trial}}}$ is decreased  while if
$E_{\hbox{\scriptsize{hub}}}(x_{\hbox{\scriptsize{trial}}})<1$, it is
increased. With this new value for $x_{\hbox{\scriptsize{trial}}}$,
$E_{\hbox{\scriptsize{hub}}}(x_{\hbox{\scriptsize{trial}}})$ is again
calculated, and the process is repeated until sufficient
accuracy is achieved. This final 
$x_{\hbox{\scriptsize{trial}}}$ is then set equal to
$x_U^{\hbox{\scriptsize{hub}}}$. A similar process is used to
determine $x_L^{\hbox{\scriptsize{disk}}}$. 

\begin{figure}
  \includegraphics[width=\linewidth]{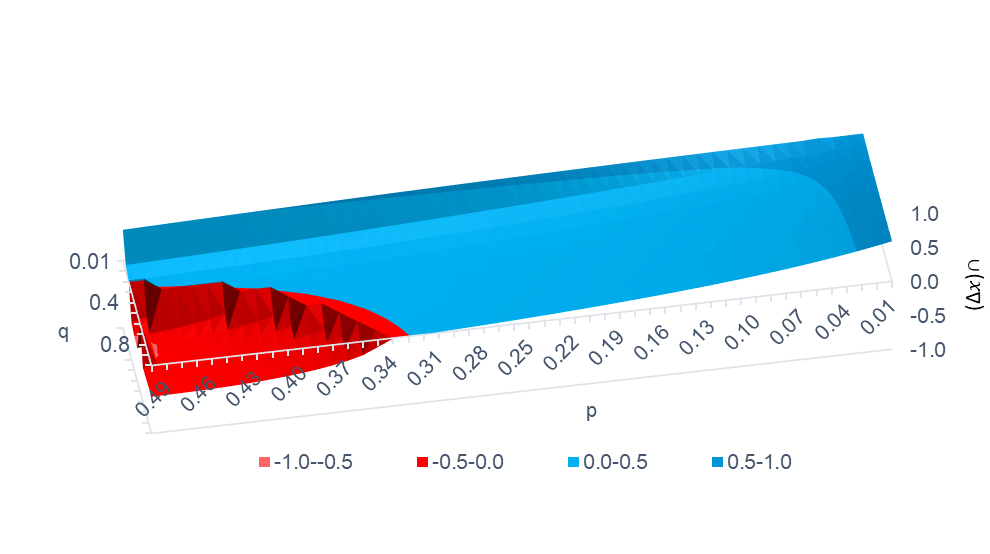}
  \caption{A 3D plot of $(\Delta x)_\cap$ with respect to $q$ and
    $p$. Note the region in red where $(\Delta
    x)_\cap<0$. In this region
    $\mathcal{R}_{\hbox{\scriptsize{out}}}\cap\mathcal{R}_{\hbox{\scriptsize{in}}}=\emptyset$.} 
  \label{Delta-X}
\end{figure}

As $(\Delta x)_\cap\equiv
x_U^{\hbox{\scriptsize{hub}}}-x_L^{\hbox{\scriptsize{disk}}}>0$ in
$\mathcal{P}_\cap$, we can use $(\Delta x)_\cap$ to determine
$\mathcal{P}_\cap$. The results of this calculation is shown in
Fig.~$\ref{Delta-X}$. Both 
$x_U^{\hbox{\scriptsize{hub}}}$ and $x_L^{\hbox{\scriptsize{disk}}}$ were calculated 
to an accuracy of $0.0001$ starting at $q=0.01$ and continuing in
$0.1$ increments from $q=0.1$ to $q=1.0$. Similarly,
$p$ starts at $0.01$, and increases in increments of $0.01$ until $p=0.50$ is
reached. We find that $(\Delta x)_\cap>0$ everywhere except for the red
triangular-shaped region shown in the figure. This region is
bounded by the lines $q=1.0$, $p=0.50$, and a curve that starts at
$(1.0,0.33)$ and ends at $(0.3, 0.50)$. Outside
of this triangular region the regions
$\mathcal{R}_{\hbox{\scriptsize{hub}}}$ and 
$\mathcal{R}_{\hbox{\scriptsize{disk}}}$ overlap, and 
there is a consistent boundary-layer solution to Eq.~$(\ref{e16})$.

We emphasize that while $(\Delta x)_\cap<0$ in the triangular-shaped region,
this does not mean that there are no solutions to Eq.~$(\ref{e16})$
in this region of $\mathcal{P}$. All that can be
concluded is that the singular perturbation analysis that divides
space into only two regions cannot be applied. A consistent solution
may be possible when a third, 
intermediate region is introduced to interpolate between the two
regions, for example. This region would be given by the solution to 
Eq.~$(\ref{e16})$ obtained by setting the inhomogeneous
term equal to the term proportional to $\epsilon^2$. However, such
solutions would depend more on the detail behavior of
$v_{\infty}(r)$ in the transition region between the galactic hub
and the disk\textemdash and thus on how this transition is
modeled\textemdash than on the asymptotic properties of the
RVC. Moreover, we have found values of $q$ and $p$ do that minimize
the stationary action, and they lie far outside of the triangular
region. As our focus is on the asymptotic behavior of RVCs, the
two-region, boundary-layer solution is sufficient for our purposes.  

The values of $(\Delta x)_\cap$ in $\mathcal{P}_\cap$ range
from $0.0014$ to $0.6339$, and, given that $6(\Delta x)_\cap <
(x_U^{\hbox{\scriptsize{hub}}}+ 
x_L^{\hbox{\scriptsize{disk}}})/2$, are quite small when compared to
either $x_U^{\hbox{\scriptsize{hub}}}$ or
$x_L^{\hbox{\scriptsize{disk}}}$. Consequently, while
$\rho_{\infty}^{\hbox{\scriptsize{disk}}}(x)$ may depend 
on the choice of $x_0\in (x_L^{\hbox{\scriptsize{disk}}},
x_U^{\hbox{\scriptsize{hub}}})$, the size of
$(\Delta x)_\cap$ is such that this choice of $x_0$ does not have much
of an impact on our analysis. Nevertheless, since we will find in
the next section that the action is dominated by the behavior of
$v_{\infty}(x)$ in the galactic disk, we choose $x_0 =
x_U^{\hbox{\scriptsize{hub}}}$ to maximize the contribution of the
galactic hub to the action. 

\section{A Spectrum of Rotational Velocity
  Curves}
\label{Spec}

In the region $\mathcal{P}_\cap$ the
boundary-layer method gives
\begin{equation}
  \rho_{\infty}(x) =
  \rho_{\infty}^{\hbox{\scriptsize{hub}}}(x)\theta(x_0-x) +
  \rho_{\infty}^{\hbox{\scriptsize{disk}}}(x)\theta(x-x_0),
\end{equation}
as the solution of Eq.~$(\ref{e16})$. Such a solution must also minimize
the action $S_{\infty}$, however. In this section we will
determine the values of $q$ and $p$ that do, and in doing so,
determine the RVCs that can form under the extended GEOM. We will find
that a continuous range of RVCs is possible, and this spectrum of RVCs is
in agreement with the URC.   

When evaluated at $L_{\infty}$ and $\rho_{\infty}$, the action
breaks up into two pieces, 
\begin{equation}
  S_{\infty}\big\vert_{(\rho_{\infty}; L_{\infty})} =
S_{\infty}^{\hbox{\scriptsize{hub}}}\big\vert_{(\rho_{\infty}^{\hbox{\scriptsize{hub}}};
  L_{\infty})} +
S_{\infty}^{\hbox{\scriptsize{disk}}}\big\vert_{(\rho_{\infty}^{\hbox{\scriptsize{disk}}}; 
  L_{\infty})},
\end{equation}
corresponding to the solutions in the regions
$\mathcal{R}_{\hbox{\scriptsize{hub}}}$ and
$\mathcal{R}_{\hbox{\scriptsize{disk}}}$. We begin with the region
$\mathcal{R}_{\hbox{\scriptsize{hub}}}$.   

\subsection{The Action in the Region $\mathcal{R}_{\hbox{\scriptsize{hub}}}$}

As $\Upsilon=\Upsilon_0+\epsilon^2 \Upsilon_1$ in the region $\mathcal{R}_{\hbox{\scriptsize{hub}}}$,
we first expand Eq.~$(\ref{e17})$ about $\Upsilon_0$, 
\begin{eqnarray}
  S_{\infty}^{\hbox{\scriptsize{hub}}}\equiv &{}&-4\pi {r_H^{*}}^3 \mathcal{E}_H \int_0^{x_0}\Bigg\{
    F(x)\Upsilon_0 +
  \frac{\alpha_{{}_\Lambda}}{1-\alpha_{{}_\Lambda}}\Upsilon_0^{1-1/\alpha_{{}_\Lambda}}
  +
  \nonumber
  \\
  &{}&
  \epsilon^2\left[\frac{1}{2}\left(\frac{d\Upsilon_0}{dx}\right)^2 +\left[F(x)-\Upsilon_0^{-1/\alpha_{{}_\Lambda}}\right]\Upsilon_1\right]
    \Bigg\} x^2dx,
\end{eqnarray}
keeping terms linear in $\Upsilon_1$ in the integrand. Here,
$\mathcal{E}_H\equiv \rho_H^{*} c^2
\mathfrak{D}_{\infty}(8\pi\rho_{H}^{*}/\Lambda_{DE})/4$. After evaluating
$S_{\infty}^{\hbox{\scriptsize{hub}}}$ at the 
solution Eq.~$(\ref{e27})$ we obtain
\begin{equation}
  \frac{S_{\infty}^{\hbox{\scriptsize{hub}}}}{4\pi
  {r_H^{*}}^3\mathcal{E}_H}\bigg\vert_{\left(\rho_{\infty}^{\hbox{\scriptsize{hub}}};
    L_{\infty}\right)} = -\int_0^{x_0}\Bigg\{\frac{F^{1-\alpha_{{}_\Lambda}}}{1-\alpha_{\Lambda}}+
  \frac{1}{2}\epsilon^2\alpha_{{}_\Lambda}^2F^{-2(\alpha_{{}_\Lambda}+1)}\left(\frac{dF}{dx}\right)^2\Bigg\}x^2dx.
\end{equation}
Importantly, in the $x\to0$ limit the first term in the integrand is
proportional to $x^{2[(\alpha_{{}_\Lambda}-1)(1-q)+1]}$ while the
second term is proportional to $x^{4\alpha_{{}_\Lambda}(1-q)}$. Since the
integral is well-defined as long as
$2[(\alpha_{{}_\Lambda}-1)(1-q)+1]>-1$ and
$4\alpha_{{}_\Lambda}(1-q)>-1$, we find that 
\begin{equation}
  q<1+\frac{1}{4\alpha_{{}_\Lambda}}.
\end{equation}
This condition is satisfied for all points in $\mathcal{P}$. 

\subsection{The Action in the Region $\mathcal{R}_{\hbox{\scriptsize{disk}}}$}
\label{A-Inner}

While in \textbf{Sec \ref{Inner}} we used $\bar{y}$ and $\bar{x}$, in
this section we find it more convenient to use 
$y=\left(\Lambda_{DE}/8\pi\rho_H^{*}\right)\bar{y}$ and $x$. Then
\begin{equation}
  y_a = \left(\frac{\Lambda_{DE}}{8\pi\rho_H^{*}}\right)\bar{y}_a,
  \quad \hbox{while}\quad
  y_1=
  \bar{x}_H^{2p}\left(\frac{v_H^{*}}{c}\right)^2\left(\frac{\Lambda_{DE}}{8\pi\rho_H^{*}}\right)\bar{y}_1. 
\end{equation}
Expanding Eq.~$(\ref{e17})$ about $\bar{y}_a$, and keeping terms linear in $y_1$,  
\begin{eqnarray}
  S_{\infty}^{\hbox{\scriptsize{disk}}}=&{}&-4\pi {r_H^{*}}^3\mathcal{E}_H\int_0^{x_{{}_{II}}}\Bigg\{
    \frac{1}{2}\epsilon^2\left(\frac{dy_a^{-\alpha_{{}_\Lambda}}}{dx\>\>\>}\right)^2+
  \frac{\alpha_{{}_\Lambda}}{1-\alpha_{{}_\Lambda}}y_a^{1-\alpha_{{}_\Lambda}}-
  \nonumber
  \\
  &{}&
  \alpha_{{}_\Lambda}\epsilon^2\frac{dy_a^{-\alpha_{{}_\Lambda}}}{dx\>\>\>}
  \frac{d\>\>\>}{dx}\left(y_a^{-(1+\alpha_{{}_\Lambda})}y_1\right)+[F(x)+\alpha_{{}_\Lambda}y_1]y^{-\alpha_{{}_\Lambda}}_a
  \Bigg\}x^2dx.
  \label{e33}
\end{eqnarray}
This action naturally breaks up into two additional pieces,
$S_{\infty}^{\hbox{\scriptsize{disk}}}=
S_{\infty}^{\hbox{\scriptsize{disk-asym}}}+S_{\infty}^{\hbox{\scriptsize{disk-near}}}$,
with the first piece consisting of the first two terms in
Eq.~$(\ref{e33})$. When $S_{\infty}^{\hbox{\scriptsize{disk-asym}}}$
is evaluated at the solution Eq.~$(\ref{e30})$, they can be integrated to give
\begin{eqnarray}
  \frac{S_{\infty}^{\hbox{\scriptsize{disk-asym}}}}{4\pi {r_H^{*}}^3\mathcal{E}_H}\bigg\vert_{\left(\rho_{\infty}^{\hbox{\scriptsize{disk}}},L_{\infty}\right)}
  =-\frac{2\alpha_{{}_\Lambda}^2(\alpha_{{}_\Lambda}\epsilon^2)^{\frac{1-\alpha_{{}_\Lambda}}{1+\alpha_{{}_\Lambda}}}(1+\alpha_{{}_\Lambda})^3\Sigma^2}{(\alpha_{{}_\Lambda}-1)(1+3\alpha_{{}_\Lambda})^2(1+5\alpha_{{}_\Lambda})}
  \left[x_{{}_{II}}^{\frac{1+5\alpha_{{}_\Lambda}}{1+\alpha_{{}_\Lambda}}}-x_0^{\frac{1+5\alpha_{{}_\Lambda}}{1+\alpha_{{}_\Lambda}}}\right],
\end{eqnarray}
after Eq.~$(\ref{e30b})$ is used. For the second piece consisting of
the third and fourth terms in Eq.~$(\ref{e33})$, after an integration
by parts and making use of Eq.~$(\ref{e31})$, it reduces to 
\begin{eqnarray}
  \frac{S_{\infty}^{\hbox{\scriptsize{disk-near}}}}{4\pi {r_H^{*}}^3\mathcal{E}_H}\bigg\vert_{\left(\rho_{\infty}^{\hbox{\scriptsize{disk}}},L_{\infty}\right)}=
  &{}&\frac{1}{2}\frac{(1+\alpha_{{}_\Lambda})^2}{(1+3\alpha_{{}_\Lambda})}\Sigma\left(\alpha\epsilon^2\right)^{-\frac{\alpha_{{}_\Lambda}}{1+\alpha_{{}_\Lambda}}}
  \nonumber
  \\
  &{}&
  \bigg\{
  \frac{\alpha_{{}_\Lambda}(1+\alpha_{{}_\Lambda})}{(1+3\alpha_{{}_\Lambda})}\bigg[x_{{}_{II}}^{\frac{1+3\alpha_{{}_\Lambda}}{1+\alpha_{{}_\Lambda}}}\left(x^2y_1\right)\Big\vert_{x_{{}_{II}}}-
    \nonumber
    \\
    &{}&x_0^{\frac{1+3\alpha_{{}_\Lambda}}{1+\alpha_{{}_\Lambda}}}\left(x^2y_1\right)\Big\vert_{x_0}\bigg]+
  \nonumber
  \\
  &{}&
  \frac{1}{3}
  \int_{x_0}^{x_{{}_{II}}}x^{\frac{2\alpha_{{}_\Lambda}}{1+\alpha_{{}_\Lambda}}}
  \frac{d\left(x\widehat{v}_{\infty}^2\right)}{dx\>\>\>}dx\bigg\}.
  \label{e33b}
\end{eqnarray}
Making use of Eq.~$(\ref{e31b})$ again, this last integral
becomes for $x_c<x_0$,
\begin{eqnarray}
\frac{1}{3}
  \int_{x_0}^{x_{{}_{II}}}x^{\frac{2\alpha_{{}_\Lambda}}{1+\alpha_{{}_\Lambda}}}
  \frac{d\left(x\widehat{v}_{\infty}^2\right)}{dx\>\>\>}dx
  =
  &{}&
  \frac{1}{3}(1+p/q)x_c^{\frac{1+3\alpha_{{}_\Lambda}}{1+\alpha_{{}_\Lambda}}-2p}\sum_{n=0}^\infty\frac{(-1)^n(1-Z_{pq}^n)}{\frac{1+3\alpha_{{}_\Lambda}}{1+\alpha_{{}_\Lambda}}-Z_{pq}^n} 
  \nonumber
  \\
  &{}&
  \bigg\{
  \left(\frac{x_c}{x_{{}_{II}}}\right)^{Z_{pq}^n-\frac{1+3\alpha_{{}_\Lambda}}{1+\alpha_{{}_\Lambda}}}
  -
  \left(\frac{x_c}{x_0}\right)^{Z_{pq}^n-\frac{1+3\alpha_{{}_\Lambda}}{1+\alpha_{{}_\Lambda}}}
  \bigg\},
  \label{e70}
\end{eqnarray}
while for $x_c>x_0$,
\begin{eqnarray}
\frac{1}{3}
  \int_{x_0}^{x_{{}_{II}}}x^{\frac{2\alpha_{{}_\Lambda}}{1+\alpha_{{}_\Lambda}}}
  \frac{d\left(x\widehat{v}_{\infty}^2\right)}{dx\>\>\>}dx
  =
  &{}&
  \frac{1}{3}(1+p/q)x_c^{\frac{1+3\alpha_{{}_\Lambda}}{1+\alpha_{{}_\Lambda}}-2p}\sum_{n=0}^\infty
  (-1)^n
  \nonumber
  \\
  &{}&
  \Bigg\{\frac{1+Z_{qp}^n}{\frac{1+3\alpha_{{}_\Lambda}}{1+\alpha_{{}_\Lambda}}+Z_{qp}^n}
  \left[1-\left(\frac{x_0}{x_c}\right)^{Z_{qp}^n+\frac{1+3\alpha_{{}_\Lambda}}{1+\alpha_{{}_\Lambda}}}\right]+
  \nonumber
  \\
  &{}&
    \frac{1-Z_{pq}^n}{\frac{1+3\alpha_{{}_\Lambda}}{1+\alpha_{{}_\Lambda}}-Z_{pq}^n}
    \left[\left(\frac{x_c}{x_0}\right)^{Z_{qp}^n-\frac{1+3\alpha_{{}_\Lambda}}{1+\alpha_{{}_\Lambda}}}-1\right]
    \Bigg\}.
    \label{e71}
\end{eqnarray}

\subsection{Minimization of
  $S_{\infty}\big\vert_{(\rho_{\infty},L_{\infty})}$}
\label{P-q-curve}

When evaluated at $\rho_{\infty}$ the action becomes
\begin{equation}
  S_{\infty}\big\vert_{\left(\rho_{\infty},L_{\infty}\right)}=
  S_{\infty}^{\hbox{\scriptsize{hub}}}\big\vert_{\left(\rho_{\infty}^{\hbox{\scriptsize{hub}}},L_{\infty}\right)}+
  S_{\infty}^{\hbox{\scriptsize{disk-asym}}}\big\vert_{\left(\rho_{\infty}^{\hbox{\scriptsize{disk}}},L_{\infty}\right)}+
  S_{\infty}^{\hbox{\scriptsize{disk-near}}}\big\vert_{\left(\rho_{\infty}^{\hbox{\scriptsize{disk}}},L_{\infty}\right)},
\end{equation}
The first two terms depend on $q$ and $p$ indirectly, through
$x_0$. Since the integrand in the first term is integrable, and because we chose
$x_0=x_U^{\hbox{\scriptsize{hub}}}\sim 1-5$, they do not 
contribute appreciably to the action. It is the third term, with its
dependence on $x_{{}_{II}}=2.264\times 10^5$, $q$, and $p$, that dominates the
behavior of $S_{\infty}\big\vert_{(\rho_{\infty},L_{\infty})}$, and will
determine the values of $q$ and $p$ that minimizes it. To emphasize
this, and to isolate the dependence of
$S_{\infty}\big\vert_{(\rho_{\infty},L_{\infty})}$
on these parameters, we define  
\begin{equation}
  S(q,p) \equiv \frac{2(1+3\alpha_{{}_\Lambda})^2}{\alpha_{{}_\Lambda}(1+\alpha_{{}_\Lambda})^3\Sigma}
  \frac{(\alpha_{{}_\Lambda}\epsilon^2)^{\frac{\alpha_{{}_\Lambda}}{1+\alpha_{{}_\Lambda}}}}
  {x_{{}_{II}}^{(1+3\alpha_{{}_\Lambda})/(1+\alpha_{{}_{\Lambda}})}}
    \frac{{S}_{\infty}^{\hbox{\scriptsize{disk-near}}}}{4\pi r_H^{*}\mathcal{E}_H}
  \bigg\vert_{\left(\rho_{\infty},L_{\infty}\right)}.
\end{equation}
Note that the dependence of $S(q,p)$ on $\mathcal{P}_\cap$ is dominated by the
$x_{{}_{II}}^2y_1(x_{{}_{II}})$ term in Eq.~$(\ref{e33b})$. This $x_{{}_{II}}^2y_1(x_{{}_{II}})$
in turn is dominated by two terms, one coming from the
particular solution to Eq.~$(\ref{e31})$, which is proportional to
$(x_c/x_{{}_{II}})^{Z_{pq}^0}$, and the other coming from the homogeneous solution
to Eq.~$(\ref{e31})$, which is proportional to
$(x_c/x_{{}_{II}})^{1/2}$. For
$S(q,p)$ to be small, the homogeneous solution must dominate, and
thus it is in the region of $\mathcal{P}_\cap$ where $Z_{pq}^0\ge
1/2$\textemdash which in turn requires $p\ge 1/4$\textemdash that the
minima of $S(q,p)$ will be found.  

The precise values for $q$ and $p$ that minimizes $S(q,p)$ are
determined numerically using the following process. We first determine 
$x_U^{\hbox{\scriptsize{hub}}}$ to an accuracy of $10^{-6}$ for $q$
starting at $0.01$ and continuing in $0.1$ increments from $0.1$ to
$1.0$, and for $p$ starting at $0.01$ and continuing in $0.01$
increments to $0.50$. We then set $x_0=x_U^{\hbox{\scriptsize{hub}}}$,
and calculate $S(q,p)$ for these values of $q$ and $p$. The result is a 
two-dimensional surface above $\mathcal{P}_\cap$. We find that there is a
slight concavity in the surface in the rectangular region of
$\mathcal{P}_\cap$ bounded by the lines $q=0.01, q=0.40, p=0.34,
p=0.50$. For each choice of $q$ in this region there is a $p(q)$ such
that  
\begin{equation}
  \frac{\partial S}{\partial p}\bigg\vert_{\left(q,p(q)\right)}=0.
  \label{e34}
\end{equation}
The solution to Eq.~$(\ref{e34})$ gives $p(q)$ as a function of $q$, and
thus defines a curve on the base space $\mathcal{P}_\cap$. The lifting 
of this curve to the surface $S(q,p)$ gives the curve
$\mathbf{S}(q)=\left(q,p(q), S(q,p(q))\right)$ on
which the action is local minimum for each choice of $q$.

\begin{figure}
  \includegraphics[width=\linewidth]{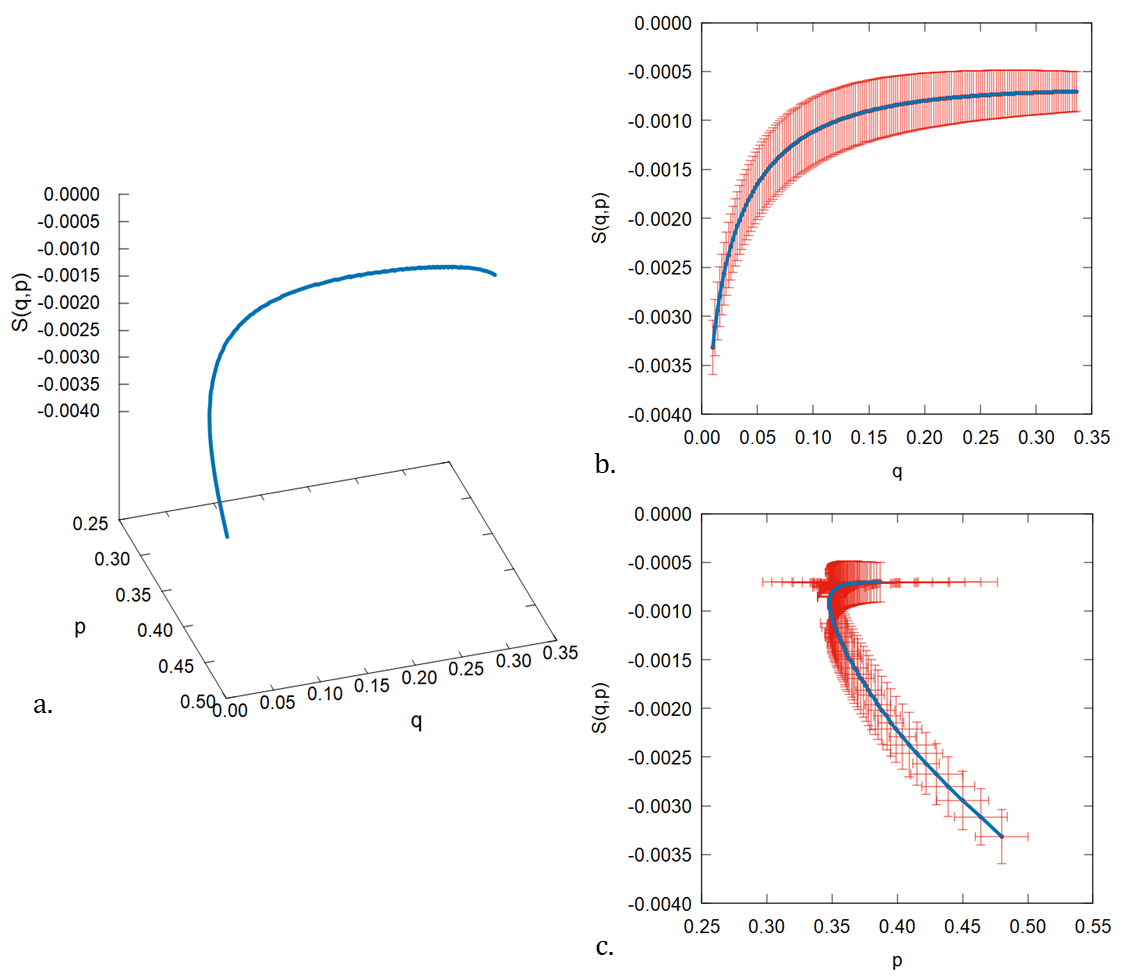}
  \caption{A series of plots of curve $\mathbf{S}(q)$. Here figure a gives
    the full three-dimensional plot of the curve, while 
    figure b gives the projection of the curve onto the $S(q,p)-q$ plane
    and figure c gives its projection onto the $S(q,p)-p$ plane. Notice
  that in both projections the minimum action curve approaches an
  asymptote.}
  \label{Action-Comb}
\end{figure}

\begin{figure}
  \includegraphics[width=\linewidth]{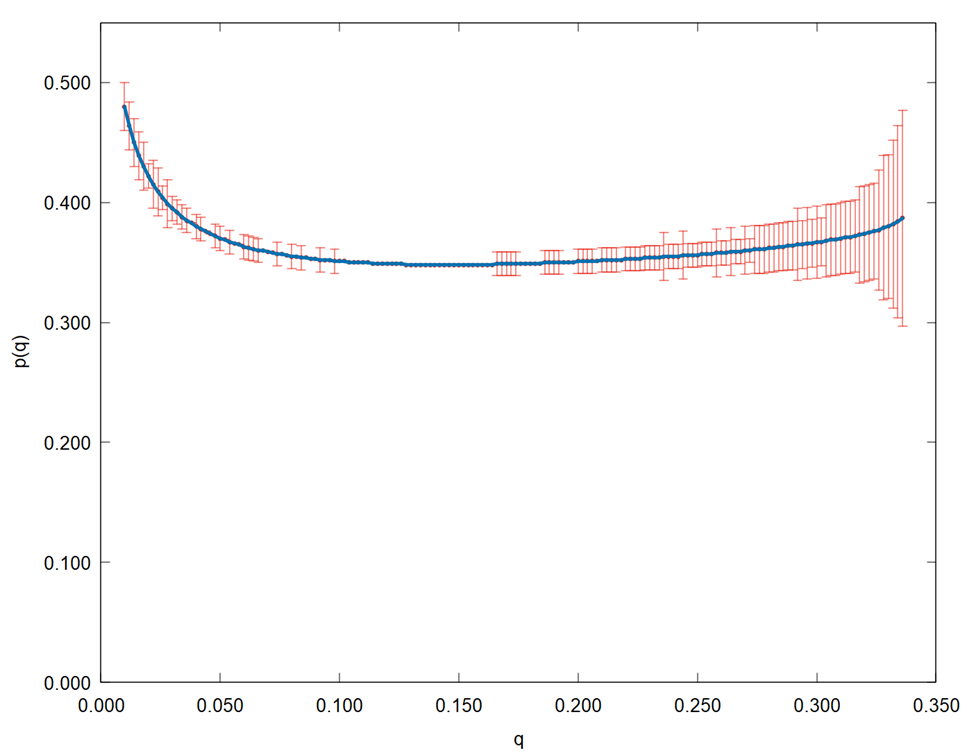}
  \caption{Graph of the dependence of $p(q)$ on $q$ along
    $\mathbf{S}(q)$. Notice that for much of the graph $p(q)$ is a weak 
    function of $q$. Notice also that the uncertainty in $p(q)$ grows rapidly
    as $q\to0.336$. } 
  \label{P-q}
\end{figure}

To determine this function $p(q)$ and the curve $\mathbf{S}(q)$, we determine
$x_U^{\hbox{\scriptsize{hub}}}$ to an accuracy of $10^{-6}$ for $q$ starting
at $0.200$ and continuing in $0.002$ increments until $0.210$ is
reached, and for $p$ starting at $0.345$ and continuing in $0.001$
increments until $0.355$ is reached. These
$x_U^{\hbox{\scriptsize{hub}}}$ are then used to calculate $S(q, p)$,
and for each $q$ the $p$ that minimizes $S(q,p)$ is determined along
with the value of $S(q,p)$ at this point. Up to $N=15000$ terms in the
series in Eqs. $(\ref{e32b})-(\ref{e32d})$ and $(\ref{e70})-(\ref{e71})$
is used in calculating $S(q,p)$. As expected, the collection of these
points form a curve on the action surface $S(q,p)$. We then follow
this curve along values of $q$ that are less than $0.200$ and along values
of $q$ that are greater than $0.210$ in increments of 
$0.002$ until we reach a $q\in\mathcal{P}_\cap$ for which a minimum of the
action cannot be found. The result of this calculation is shown in
Figs.~$\ref{Action-Comb}$ and $\ref{P-q}$. 

Figure $\ref{Action-Comb}$a is a graph of the minimum action curve
$\mathbf{S}(q)$ above a region of the base parameter space
$\mathcal{P}_\cap$; not shown is the surface on which this curve
lies. The projection of this curve onto the $S(q,p)-q$ plane is shown
in Fig.~$\ref{Action-Comb}$b, while the projection of the curve onto
the $S(q,p)-p$ plane is given in Fig.~$\ref{Action-Comb}$c. Notice the
asymptote for the curve shown in Fig.~$\ref{Action-Comb}$b and
$\ref{Action-Comb}$c.   

\begin{figure}
  \includegraphics[width=\linewidth]{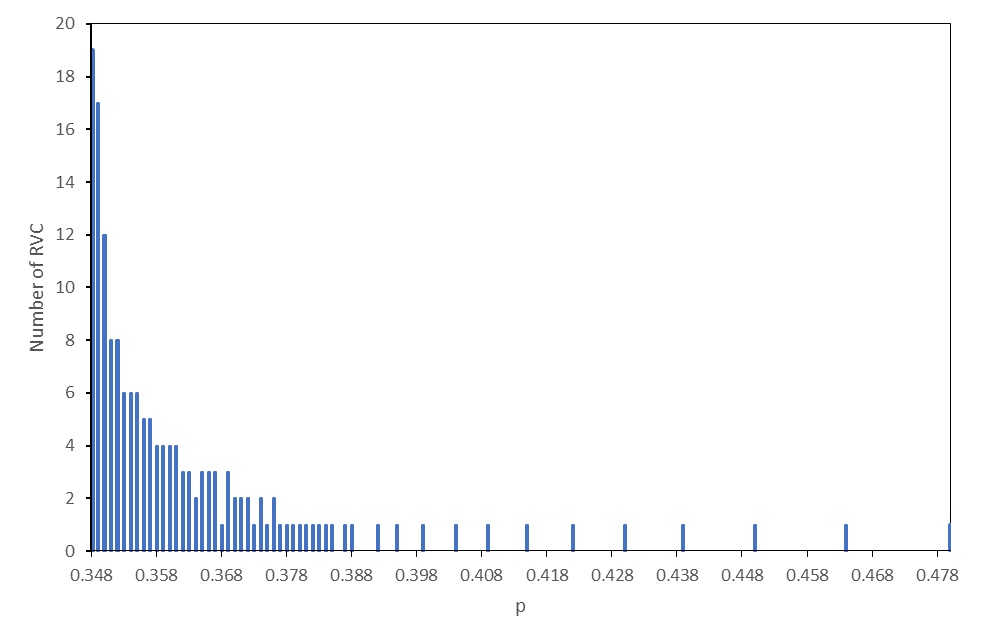}
  \caption{Histogram showing the distribution of $p$ with the most
    probable value of $p$ being $0.348$. } 
  \label{Hist}
\end{figure}

Figure $\ref{P-q}$ shows the graph of $p(q)$ versus $q$, and is
the projection of $\mathbf{S}(q)$ onto the $q-p$ plane. Each
point $(q,p(q))$ on the curve gives a $L_{\infty}(x)$ that results
in a density $\rho_{\infty}$ that minimizes 
$S_{\infty}$. Thus, each point $(q,p(q))$ on this curve
gives the density and, through $v_{\infty}(x)$, the RVC of a galaxy
that can form under the extended GEOM. Notice that $p(q)$ is
nearly flat for most of the values of $q$ shown, and thus many of these
galaxies will have RVCs that have similar asymptotic 
behavior in the galactic disk, while at the same time have
very different behavior in the galactic hub. This can be seen
explicitly in Fig.~$\ref{Hist}$. 

By focusing only on the large $x$ asymptotic behavior of the RVC, we
use the increments 0.001 by which $p$ was increased when
determining $\mathbf{S}(q)$ as a bin size $\Delta p$, and determine
the probability of finding a galaxy with an 
asymptotic power-law exponent between $p$ and $p+\Delta p$. This is
done by simply counting the number of RVCs with a value for $p$
between $p$ and $p+\Delta p$ without regard to their 
corresponding values of $q$; the resultant histogram is shown in  
Fig.~$\ref{Hist}$. Notice that the most probable value of
$p$\textemdash with 19 out of the 164 possible RVCs, or
11.6\%\textemdash is $0.348$. The median of this distribution of
curves is at $p=0.349_{\pm 0.01}$; the corresponding RVC has a 
$(q,p)=(0.172, 0.349_{\pm 0.01})$. In addition, $50\%$ of the possible RVCs
have a $p\le 0.355$ while $95\%$ of the curves have a $p\le 0.404$.

Shown also in Figs.~$\ref{Action-Comb}$ and $\ref{P-q}$ are the
estimated uncertainties in determining $S(q,p(q))$ and $p(q)$. Notice in
particular the large increase in uncertainty in $p(q)$ as 
$q\to 0.336$; this is precisely the location of the asymptote for
$\mathbf{S}(q)$. The largest contribution to the uncertainties is due to
the power-law exponent $\alpha_{{}_\Lambda}$. While the uncertainties in
$r_H^{*}, v_H^{*}$, and $\lambda_{DE}$ also contribute to the
uncertainties in $S(q,p(q))$ and $p(q)$, including these contributions
to the uncertainties in any reliable manner would require determining  
$x_U^{\hbox{\scriptsize{hub}}}$ to an accuracy much higher than
$10^{-6}$; we did not do so here. Instead, we focus on the uncertainty
due to $\alpha_{{}_\Lambda}$ by increasing the 
value of $\alpha_{{}_\Lambda}$ to $\alpha_{{}_\Lambda}+\Delta
\alpha_{{}_\Lambda}$ with $\Delta\alpha_{{}_\Lambda}= 0.01$. A new
curve $\mathbf{S}(q)$ was then calculated, and the graph $p(q)$
determined. The uncertainty in $S(q,p(q))$ was calculated from the
difference in $S(q,p(q))$ due to this change in
$\alpha_{{}_\Lambda}$; the uncertainty in $p(q)$ was calculated in a 
similar way. While large, this is the smallest 
$\Delta\alpha_{{}_\Lambda}$ that could be used without increasing the 
accuracy in $x_U^{\hbox{\scriptsize{hub}}}$ to beyond $10^{-6}$. For
these reasons we caution that the uncertainty shown in
Figs.~$\ref{Action-Comb}$ and $\ref{P-q}$ is an estimate.   

After analyzing a homogeneous sample of 1100 RVCs, Persic
et.~al.~\cite{Pers1996} found that the profile of the RVC for a 
galaxy is determined by a single parameter, the luminosity of the
galaxy. They further showed that these profiles could be described by
a single function of this luminosity. This work 
was further refined by Salucci et.~al.~\cite{Salu2007} where
they expressed the square of the URC as the sum of two terms,
$V_{URC}^2=V_{URCD}^2+V_{URCH}^2$. This $V_{URCD}$ gives the
stellar contribution to the URC, while $V_{URCH}$ gives the dark matter
component. To demonstrate the self-similarity of the URC, and to
compare the URC to $V_{NFW}$, an ensemble of URCs, each with
a different virial mass $M_{vir}$, was plotted in Fig.~4 of
\cite{Salu2007}. This was accomplished by rescaling
$x_{opt}=r/R_{opt}\to x_{vir}=r/R_{vir}$, where $R_{opt}$ and
$R_{vir}$ are the optical and virial radii, respectively, and
normalizing both the ensemble of URCs and the $V_{NFW}$ so that all
the curves agree at $x_{vir}=1$. The similarity between these curves
becomes readily apparent.  

\begin{figure}
  \includegraphics[width=\linewidth]{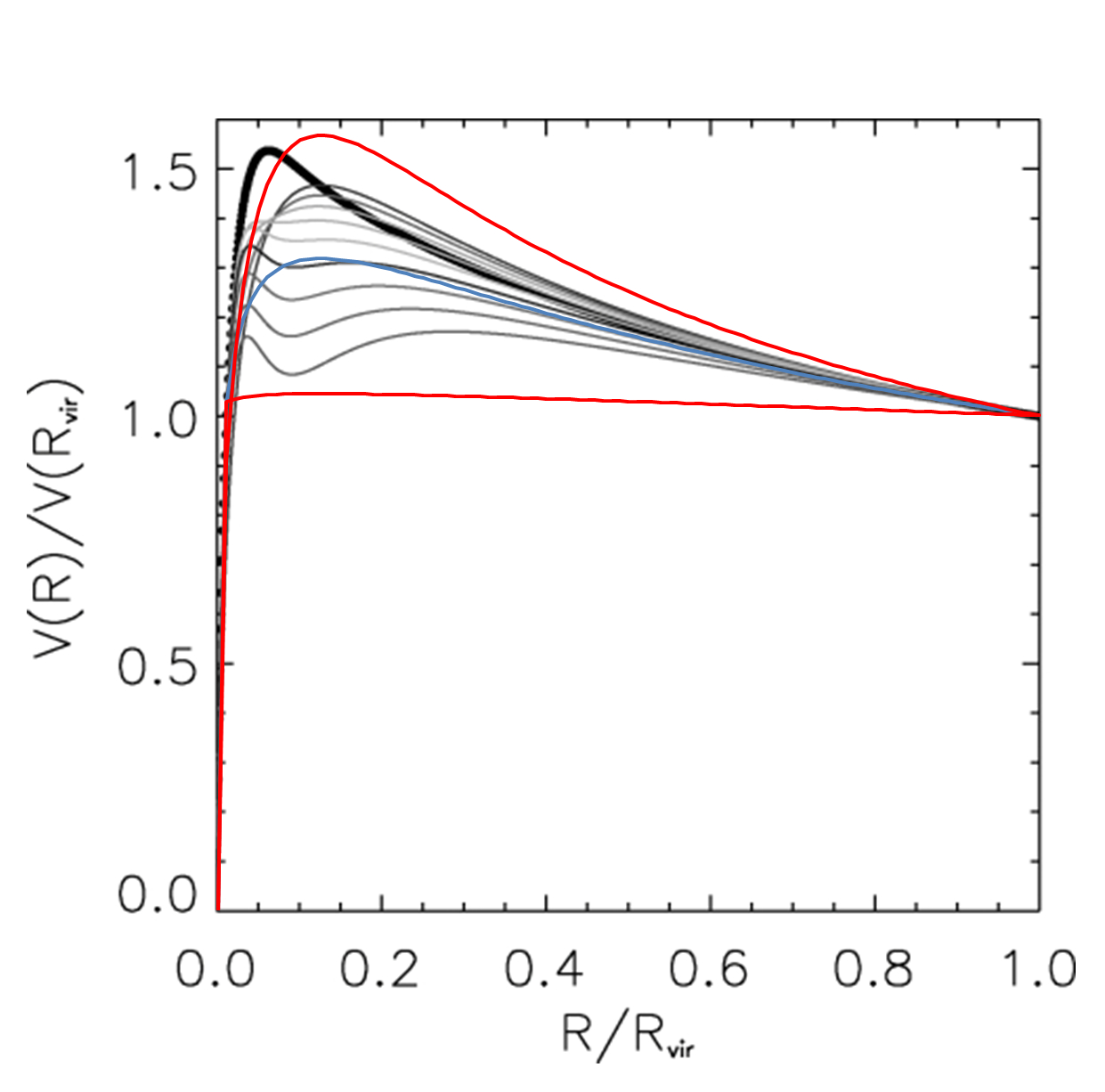}
  \caption{Graphs of the extended GEOM RVC superimposed on Fig.~4 of
    \cite{Salu2007}. The two graphs in red is given below by the $(0.010,  
    0.048_{\pm 0.020})$ curve, and above by the $(0.336. 0.387_{\pm
      0.090})$ curve. They bracket the ensemble of UHCs 
    (grey-scale lines) and $V_{NFW}$ (solid black line) from
    \cite{Salu2007}. Also plotted is the median $(0.172, 0.349_{\pm
      0.010})$ curve in blue. } 
  \label{Salu-Comp}
\end{figure}

By rescaling $x\to x_{\hbox{\scriptsize{scale}}}=x/8$ so that the
maximum of $\widehat{v}_{\infty}(x)$ now occurs near the location of the
maxima of the URCs in Fig.~4 of \cite{Salu2007}, and
rescaling $\widehat{v}_{\infty}$ so that 
$\widehat{v}_{\infty}(x_{\hbox{\scriptsize{scale}}}=1)=1$, we have 
added the spectrum of RVCs predicted by the extended GEOM to this
graph. The result is shown in Fig.~$\ref{Salu-Comp}$. The two ends of
the $p(q)$ graph in Fig.~$\ref{P-q}$ correspond to $(0.010,
0.480_{\pm0.020})$ and $(0.336, 0.387_{\pm 0.090})$; all the predicted
RVCs are bracketed below by the 
$(0.010, 0.480_{\pm0.020})$ curve and above by the $(0.336,
0.387_{\pm0.090})$ curve. These two curves, shown in red in
Fig.~$\ref{Salu-Comp}$, are superimposed on Fig.~4 of  
\cite{Salu2007} along with the median RVC curve given by $(0.172, 0.349_{\pm
  0.010})$; this median curve is shown in blue in
Fig.~$\ref{Salu-Comp}$. The two extreme RVCs, $(0.010, 0.480_{\pm{0.020}})$  
and $(0.336, 0.387_{\pm 0.090})$, also bracket the ensemble of URCs from
\cite{Salu2007}. While the $(0.336, 0.387_{\pm 0.090})$ curve lies
significantly above the highest URC curve shown, the uncertainty $p$
for this curve is both very large and is the highest of the 
extended GEOM RVCs. The median RVC curve $(0.172, 0.349_{\pm 0.010})$
of the extended GEOM lies also in the middle of the ensemble of URC
graphed in Fig.~4 of \cite{Salu2007}. While the two extreme curves
from the extended GEOM does not approach the $V_{NFW}$ RVC as closely
as the URC curves, in the region $2\le x\le 18$ Salucci et.~al.~have
shown that the Burkert and NFW profiles can be approximated as  
\begin{equation}
  V_{URCH}(x)=V_{URCH}(3.24)\frac{2.06x^{0.86}}{1.59+x^{1.19+\epsilon_{NFW}}},
\end{equation}
where $\epsilon_{NFW}<0.1$\footnote{In
  \cite{Salu2007} the variable $y$ is used for the ratio $r/r_H^{*}$
  instead of $x$.}, and for $x$ near $18$, $V_{URCH}(x)\sim
x^{-0.33-\epsilon_{NFW}}$.   The histogram in
Fig.~$\ref{Hist}$ shows that the most probable asymptotic behavior of
a RVC predicted by the extended GEOM has a $p=0.348$. Such a RVC would
have the asymptotic behavior $\widehat{v}_{\infty}\sim x^{-0.348}$,
in good agreement with the $V_{NFW}$, and the ensemble of URC
curves. While the extreme curve $(0.010, 0.480_{\pm 0.020})$ does not
have a $p$ that is within $\epsilon_{NFW}$ of $0.33$, $95\%$ of the
predicted curves have a $p\le 0.404$, and is within $\epsilon_{NFW}$
of $0.33$.   

\section{Stability Analysis}
\label{Stab}

We now turn our attention to the stability analysis of the stationary
solutions found in the previous section. The equation determining
the perturbation $u_1^r(t,r)$ of the
radial velocity in the stationary limit was found in
\textbf{Sec \ref{Perturb}}. Instead of working with $u_1^r(t,r)$,
however, we work with the radial current density 
$j^r(t,r)=\rho(t,r)u^r(t,r)\approx\rho_{\infty}(r)u_1^r(t,r)$ since
$u_{\infty}(r)=0$, and through it, the 
flux of mass through a sphere $Sph(r)$ of radius $r$,
\begin{equation}
  \mu(t,r)\equiv\int_{Sph(r)} \vec{\jmath}\cdot d\vec{A} = 4\pi
  r^2\rho_{\infty}(r)u_1^r(t,r).
\end{equation}
Equation $(\ref{e24})$ then becomes 
\begin{eqnarray}
  0=&{}& \frac{\partial^2\mu}{\partial t^2} +
  c^2\chi\left(\frac{4\pi\rho_{\infty}}{\Lambda_{DE}}\right)r^2\frac{\partial\>\>\>}{\partial
    r}
  \left[\left(\frac{\Lambda_{DE}}{8\pi\rho_{\infty}}\right)^{1+\alpha_{{}_\Lambda}}\frac{1}{r^2}\frac{\partial
      \mu}{\partial r}\right]+
  \nonumber
  \\
  &{}&
  \left[\frac{1}{r^3}\frac{\partial L_{\infty}^2}{\partial r} -4\pi
    G\rho_{\infty}\right]\mu.
\label{e35}
\end{eqnarray}
The solution of this differential equation depends on
$\rho_{\infty}$, and is different in the two regions. We begin with
the region $\mathcal{R}_{\hbox{\scriptsize{hub}}}$.

\subsection{Perturbations in the Region $\mathcal{R}_{\hbox{\scriptsize{hub}}}$}
\label{Per-Outer}

Following the notion in \textbf{Sec} $\ref{A-Inner}$, we take
$y=\rho_{\infty}^{\hbox{\scriptsize{hub}}}/\rho_H^{*}$, and Eq.~$(\ref{e35})$ becomes 
\begin{equation}
  0=\frac{1}{3{\omega_{H}^{*}}^2}\frac{\partial^2\mu}{\partial t^2}
  +\alpha_{{}_\Lambda}\epsilon^2yx^2\frac{\partial\>\>\>}{\partial
    x}\left[\frac{y^{-(1+\alpha_{{}_\Lambda})}}{x^2}\frac{\partial\mu}{\partial
      x}\right]+\left[ \frac{\widehat{v}_{\infty}^2}{3x^2}+F-y\right]\mu,
  \label{e36}
\end{equation}
where $\omega_{H}^{*}=v_H^{*}/r_H^{*}$ is the angular velocity of the
galactic hub at $r_H^{*}$. We are interested in the normal modes 
\begin{equation}
  \mu(t,x) = e^{i\sqrt{3}\omega_H^{*}
    {\varepsilon_{\hbox{\scriptsize{h}}}} t} H(x),
\end{equation}
that oscillate with frequency $\omega=\sqrt{3}
{\varepsilon_{\hbox{\scriptsize{h}}}}\omega_H^{*}$. Then by taking
\begin{equation}
  H(x) =  xy^{(1+\alpha_{{}_\Lambda})/2}\xi(x),
\end{equation}
Eq.~$(\ref{e36})$ reduces to a particularly simple form,
\begin{equation}
  0=\alpha_{{}_\Lambda}\epsilon^2\frac{d^2\xi}{dx^2}+M^2(x)\xi(x),
  \label{e37}
\end{equation}
where $M^2(x) = M^2_0(x)-(2\Gamma^2(x)/x^2)\epsilon^2$ after expanding to first
order in $\epsilon^2$. Here
\begin{eqnarray}
  M_0^2(x) &=&
  \left[\frac{\widehat{v}_{\infty}^2(x)}{3x^2}-{\varepsilon^2_{\hbox{\scriptsize{h}}}}\right]
  \left[F(x)\right]^{\alpha_{{}_\Lambda}}, \hbox{ and}
  \nonumber
  \\
  \Gamma^2(x) &=&
  1-\frac{1}{2}\left[\frac{1}{2}(\alpha_{{}_\Lambda}-1)+
    \alpha_{{}_\Lambda}\frac{M_0^2}{F^{1+\alpha_{{}_\Lambda}}}\right]\left[x^2\frac{d^2\log{F}}{dx^2\>\>\>}-2x\frac{d\log{F}}{dx}\right]+
  \nonumber
  \\
  &{}&
  \frac{1}{2}\left[\frac{1}{4}\left(\alpha_{{}_\Lambda}-1\right)^2+\alpha_{{}_\Lambda}^2\frac{M_0^2}{F^{1+\alpha_{{}_\Lambda}}}\right]
  \left(x\frac{d\log{F}}{dx\>\>\>}\right)^2.
\end{eqnarray}
Using the WKB approximation to order $\epsilon$, we find that  
\begin{equation}
  \xi(x) =
  \frac{1}{\sqrt{M_0}}\left[
    A^{\hbox{\scriptsize{hub}}}\cos(\Omega(x))+
   B^{\hbox{\scriptsize{hub}}}\sin(\Omega(x))
      \right],
\end{equation}
where
\begin{eqnarray}
  \Omega(x) =
  &-&
  \int_x^{x_0}\Bigg\{
    \frac{M_0(s)}{\sqrt{\alpha_{{}_\Lambda}}\epsilon}
    -
  \nonumber
  \\
  &{}&
  \frac{\sqrt{\alpha_{{}_\Lambda}}\epsilon}{s^2M_0(s)}\left[\Gamma^2(s)
    -\frac{1}{8}\left(s\frac{d\log{M_0}}{ds\>\>\>}\right)^2+
    \frac{1}{4}s^2\frac{d^2\log{M_0}}{ds^2\>\>\>}
      \right]
    \Bigg\}ds.
\end{eqnarray}

If $\varepsilon^2_{\hbox{\scriptsize{h}}}<0$, then $M_0^2(x)>0$, and
$\mu(t,x)$ will be an exponential function of $t$, and
an oscillatory function of $x$. The situation is more complicated if 
$\varepsilon^2_{\hbox{\scriptsize{h}}}>0$. While $\mu(t,x)$ will always be an 
oscillatory function of $t$, it will be an exponential
function of $x$ when $M_0^2(x)<0$, and an oscillatory function of $x$
when $M_0^2(x)>0$. We are interested in the case when $\mu(t,x)$ is an
oscillatory function of both $t$ and $x$, and therefore bounded. Given that in
$\mathcal{R}_{\hbox{\scriptsize{hub}}}$ we have $x<x_U^{\hbox{\scriptsize{hub}}}$, we can always
choose a $\varepsilon_{\hbox{\scriptsize{h}}}$ for which this is true. 

The function $\widehat{v}_{\infty}^2(x)/3x^2$ is monotonically decreasing
when $q\le1$. Moreover, when $q<1$,
$\widehat{v}_{\infty}^2(x)/3x^2\to\infty$ as $x\to0$. As $x<x_0$ for all
$x\in\mathcal{R}_{\hbox{\scriptsize{hub}}}$,
\begin{equation}
  M_0^2(x)\ge\left[\frac{\widehat{v}_{\infty}^2(x_0)}{3x_0^2}-
    \varepsilon^2_{\hbox{\scriptsize{h}}}\right]\left[F(x)\right]^{\alpha_{{}_\Lambda}}.
\end{equation}
Thus, to ensure that $M_0^2(x)\ge0$ in $\mathcal{R}_{\hbox{\scriptsize{hub}}}$, we limit
${\varepsilon_{\hbox{\scriptsize{h}}}}<
\varepsilon_{\hbox{\scriptsize{h}}}^{\hbox{\scriptsize{max}}}$ where
${\varepsilon_{\hbox{\scriptsize{h}}}^{\hbox{\scriptsize{max}}}}=
\widehat{v}_{\infty}(x_0)/\sqrt{3}x_0$. This  
in turn imposes an upper limit to the frequency, 
\begin{equation}
  \omega< \frac{\widehat{v}_{\infty}(x_0)}{x_0}\omega_H^{*},
\end{equation}
of oscillation for the mode. As $\xi(x)$ is then assured to be an
oscillatory function, we define 
\begin{equation}
  k(x) = \frac{d\Omega}{dr},
\end{equation}
as the local wavenumber for the oscillation with
$\lambda(x)=2\pi/k(x)$ being its corresponding wavelength. When $x\to0$, 
\begin{equation}
  k(x)r_H^{*}\approx
  \frac{1}{\sqrt{\alpha_{{}_\Lambda}}\epsilon}
  \left(\frac{1}{3}\left[1+q/p\right]\right)^{(1+\alpha_{{}_\Lambda})/2}
  \left(2q+1\right)^{\alpha_{{}_\Lambda}/2}x^{-(1+\alpha_{{}_\Lambda})(1-q)}.
\end{equation}
When, however, $x\to x_0^{-}$ and
$\varepsilon_{\hbox{\scriptsize{h}}}\to
\varepsilon^{\hbox{\scriptsize{max}}}_{\hbox{\scriptsize{h}}}$, we
may approximate $M_0^2(x)\approx
\left[\left(\varepsilon_{\hbox{\scriptsize{h}}}^{\hbox{\scriptsize{max}}}\right)^2
  - \varepsilon_{\hbox{\scriptsize{h}}}^2\right] + B(x_0-x)$, and
\begin{equation}
  k(x)r_H^{*} =\frac{M_0(x)}{\sqrt{\alpha}\epsilon} +
  \frac{5}{32}\frac{\sqrt{\alpha}\epsilon B^2}{M_0^5(x)}.
\end{equation}
Here,
\begin{equation}
  B=-\frac{dM_0^2}{dx}\Bigg\vert_{x_0}.
\end{equation}
The wavelength $\lambda(x)$ varies widely over $\mathcal{R}_{\hbox{\scriptsize{hub}}}$, with
the shortest wavelengths near $x=0$; it is here where $k(x)\to \infty$
when $q<1$. The longest wavelength is occurs
close to $x_0$, and the maximum wavelength for the stationary solutions found
in \textbf{Sec \ref{P-q-curve}} ranges from $\sim 1.5r_H^{*}$ to
$\sim 7r_H^{*}$. Given that the radius
$r_U^{\hbox{\scriptsize{hub}}}=x_U^{\hbox{\scriptsize{hub}}}r_H^{*}$
of the region $\mathcal{R}_{\hbox{\scriptsize{hub}}}$ varies
correspondingly from $\sim 2 r_H^{*}$ to $3.5r_H^{*}$,
this range of maximum wavelengths is reasonable, and expected. 

\subsection{Perturbations in the Region $\mathcal{R}_{\hbox{\scriptsize{disk}}}$}
\label{Per-Inner}

Using the variables introduced in \textbf{Sec
  \ref{Inner}}, Eq.~$(\ref{e35})$ becomes
\begin{eqnarray}
  0=&{}&\frac{2}{{\omega_H^{*}}^2}\frac{\partial^2\mu}{\partial t^2} -
  \bar{y}_a\frac{(1+E_{\hbox{\scriptsize{disk}}})}{\vert\Sigma^{1+\alpha_{{}_\Lambda}}\vert}
  \bar{x}^2\frac{\partial \>\>\>}{\partial
    \bar{x}}\left[\frac{1}{(1+E_{\hbox{\scriptsize{disk}}})^{1+\alpha_{{}_\Lambda}}}\frac{\partial
      \mu}{\partial
      \bar{x}}\right]-
  \nonumber
  \\
  &{}&
  \left[1+E_{\hbox{\scriptsize{disk}}}-
    \frac{1}{\bar{y}_a}\left(\bar{F}+\frac{2}{\chi}\frac{v^2}{c^2\bar{x}^2}\right)\right]\bar{y}_a\mu,
\label{e38}
\end{eqnarray}
in this region. We make the change in variable $\bar{x} \to
z=1/\sqrt{y_a(\bar{x})}$, and as with the previous section, look for normal
modes with a definite frequency,
\begin{equation}
\mu(t,z) =
e^{i\omega_{H}^{*}\varepsilon_{\hbox{\scriptsize{d}}}t/\sqrt{2}}H(z).
\end{equation}
Equation $(\ref{e38})$ reduces to a particularly simple form after taking
$H(z)=[z(1+E_{\hbox{\scriptsize{disk}}})]^{(1+\alpha_{{}_\Lambda})/2}\xi(z)$,
\begin{equation}
  0=z^2\frac{d^2\xi}{dz^2}+z\frac{d\xi}{dz}+
  \left[2(1+3\alpha_{{}_\Lambda})\varepsilon_{\hbox{\scriptsize{d}}}^2z^2+(1+\alpha_{{}_\Lambda})^2\nu^2
    + P(z)\right]\xi,
  \label{e38b}
\end{equation}
with
\begin{eqnarray}
  P(z)
  =&{}&2(1+3\alpha_{{}_\Lambda})\left[\alpha_{{}_\Lambda}\varepsilon_{\hbox{\scriptsize{d}}}^2z^2+
    (1+\alpha_{{}_\Lambda})\right]E_{\hbox{\scriptsize{disk}}}
  +
  \nonumber
  \\
  &{}&
  \frac{1}{2}(1+\alpha_{{}_\Lambda})\left[z^2\frac{d^2E_{\hbox{\scriptsize{disk}}}}{dz^2}
    -\alpha_{{}_\Lambda}z\frac{dE_{\hbox{\scriptsize{disk}}}}{dz}\right]+
  \nonumber
  \\
  &{}&
  \frac{2}{\chi}\frac{{v_H^{*}}^2}{c^2}\frac{(1+\alpha_{{}_\Lambda})^2}{z^{2\alpha_{{}_\Lambda}}}
  \left[\frac{1}{1+\alpha_{{}_\Lambda}}
    z\frac{d\widehat{v}_{\infty}^2}{dz\>\>}
    +2\widehat{v}^2_{\infty}\right].
\end{eqnarray}
This $P(z)\sim {v_H^{*}}^2/c^2$, and is small compared to the terms
proportional to $z^2$ and
$\nu^2$ terms in Eq.~$(\ref{e38b})$. We thus treat the $P(z)\xi$ term
as a perturbation, and solve Eq.~$(\ref{e38b})$ perturbatively
by taking $\xi = \xi_0+\xi_1$. Then for
$\bar{z}=\sqrt{2(1+3\alpha_{{}_\Lambda})}\varepsilon_{\hbox{\scriptsize{d}}}z$, 
\begin{eqnarray}
  0&=&\bar{z}^2\frac{d^2\xi_0}{d\bar{z}^2}+\bar{z}\frac{d\xi_0}{d\bar{z}}+
  \left[\bar{z}^2+(1+\alpha_{{}_\Lambda})^2\nu^2\right]\xi_0,
  \label{e39}
  \\
  -P(\bar{z})\xi_0&=&\bar{z}^2\frac{d^2\xi_1}{d\bar{z}^2}+\bar{z}\frac{d\xi_1}{d\bar{z}}+
  \left[\bar{z}^2+(1+\alpha_{{}_\Lambda})^2\nu^2\right]\xi_1.
  \label{e40}
\end{eqnarray}
These are Bessel's equations of imaginary order
$\bar{\nu}=(1+\alpha_{{}_\Lambda})\nu$. Using the same terminology and
notation in \cite{Duns1990}, we find
\begin{eqnarray}
  \xi_0(\bar{z}) &=&
  A^{\hbox{\scriptsize{disk}}}F_{i\bar{\nu}}(\bar{z})+
  B^{\hbox{\scriptsize{disk}}}G_{i\bar{\nu}}(\bar{z}),
  \\
  \xi_1(\bar{z})
  &=&\int_{\bar{z}_0}^{\bar{z}}P(\bar{s})\xi_0(\bar{s})G(\bar{s},\bar{z})\frac{d\bar{s}}{\bar{s}}.
  \label{e41}
\end{eqnarray}
Here, $\bar{z}_0\equiv
\sqrt{2(1+3\alpha_{{}_\Lambda})}\varepsilon_{\hbox{\scriptsize{d}}}/\bar{y}_a(x_0)^{1/2}$,
and  $G(\bar{s},\bar{z})$ is the Green's function,
\begin{equation}
  G(\bar{s},\bar{z})=\frac{\pi}{2}\bigg[G_{i\bar{\nu}}(\bar{s})F_{i\bar{\nu}}(\bar{z})-
    F_{i\bar{\nu}}(\bar{s})G_{i\bar{\nu}}(\bar{z})\bigg].
\end{equation}
In the limit $x\to\infty$, $\bar{z}\to\infty$, and $F_{i\bar{\nu}}(\bar{z})\sim
\cos(\bar{z}-\pi/4)/\sqrt{\bar{z}}$ while $G_{i\bar{\nu}}(\bar{z})\sim
  \cos(\bar{z}-\pi/4)/\sqrt{\bar{z}}$; $\xi_0$ thus dies off as
  $1/\sqrt{\bar{z}}$. For $\xi_1(\bar{z})$, we first note that
  $E_{\hbox{\scriptsize{hub}}}\sim \bar{y}_1/\bar{y}_a\sim
  \bar{z}^2\bar{y}_1$. Then 
  \begin{equation}
    P(\bar{s})\xi_0(\bar{s})
    \binom{F_{i\bar{\nu}}(\bar{s})}{G_{i\bar{\nu}}(\bar{s})}
        \sim
        \bar{z}^3\bar{y}_1(\bar{z}) \binom{\cos(\bar{s}-\pi/4)}{\sin(\bar{s}-\pi/4)},
  \end{equation}
  since $P(\bar{z})\sim \bar{z}^2E_{\hbox{\scriptsize{disk}}}$. From
  \textbf{Sec \ref{Inner}}, $y_1(z)$ consists of a particular solution $\bar{y}_P$ and a
  homogenous solution $\bar{y}_h$. The particular solution
  behaves as $y_p(\bar{z}) \sim
  1/\bar{z}^{2(1+\alpha_{{}_\Lambda})(1+p)}$ for large $\bar{z}$, and
  as such $\bar{z}^3\bar{y}_p\sim 1/\bar{z}^{2(1+\alpha_{\Lambda})(1+p)-3}$. For
    the integral Eq.~$(\ref{e41})$ to converge at large $\bar{z}$,
      $2(1+\alpha_{\Lambda})(1+p)-3>0$ or
      $p>(1-2\alpha_{{}_\Lambda})/2(1+\alpha_{{}_\Lambda})$. This is
      always true on $\mathcal{P}$. Next, for the homogenous solution,
      $\bar{y}_h(\bar{z})\sim 
  1/\bar{z}^{5(1+\alpha_{{}_\Lambda})/2}$, so that
  $\bar{z}^3\bar{y}_h(\bar{z})\sim
  1/\bar{z}^{(5\alpha_{{}_\Lambda}-1)/3}$. The
  contribution by $\bar{y}_h(\bar{z})$ to the integral also
  converges. Thus, $\xi(\bar{z})$ is well-behaved everywhere.

\subsection{The Stability of Stationary Solutions}

From \textbf{Sec \ref{Per-Inner}} we see that in $\mathcal{R}_{\hbox{\scriptsize{disk}}}$
small perturbations in the current flux $\mu$ will remain
small, and the stationary solutions to \textbf{\textit{Evol}} found in
\textbf{Sec \ref{P-q-curve}} are thus stable. The
situation is more complicated in $\mathcal{R}_{\hbox{\scriptsize{hub}}}$, however.

We see from the analysis in \textbf{Sec \ref{Per-Outer}} that $\mu$ be
an oscillatory\textemdash and thus bounded\textemdash 
function of both $t$ and $r$ when
$\varepsilon_{\hbox{\scriptsize{h}}}^2\ge 0$ and $0\le 
\varepsilon_{\hbox{\scriptsize{h}}}<\varepsilon_{\hbox{\scriptsize{h}}}^{\hbox{\scriptsize{max}}}$. This
only occurs when the frequency of oscillations of the perturbation $\omega<
\omega_H^{*}\widehat{v}_{\infty}(x_0)/x_0$. For 
the stationary solutions found in \textbf{Sec \ref{P-q-curve}},
$0.267_{\pm 0.076} \le \widehat{v}_{\infty}(x_0)/x_0 \le
0.47_{\pm0.16}$. These stationary solutions are 
therefore stable in $\mathcal{R}_{\hbox{\scriptsize{hub}}}$ as long as the period of
oscillations for the perturbations is longer than
$T_{\hbox{\scriptsize{max}}}$ with $0.91_{\pm0.31} \le
T_{\hbox{\scriptsize{max}}}\le 1.58_{\pm 0.46}$ billion years. Such
perturbations have a maximum wavelength of $\sim 1.5 r_H^{*}$ to $\sim
7 r_H^{*}$. 

\section{Concluding Remarks}
\label{End}

The choice of $v_{\infty}(x)$; the direct connection between
the parameters used in its construction and observations; and the
ability of $v_{\infty}(x)$ to model a wide range of rotational
velocity profiles, have allowed for those profiles that are consistent
with the extended GEOM to be 
determined. Indeed, while each point in $\mathcal{P}_\cap$ corresponds to a
different RVC, and while each RVC may give a $L_{\infty}(r)$ that
results in a solution $\rho_{\infty}(r)$ of the stationary \textbf{\textit{Evol}}, it is
only along the curve $(q,p(q))$ in $\mathcal{P}_\cap$ for which a stationary
solution that minimizes the action can be found. As each $\rho_{\infty}(r)$
obtained from the stationary \textbf{\textit{Evol}} would correspond a
galaxy with a RVC given by $v_{\infty}(x)$, it is therefore only 
galaxies with RVCs given along this curve that will be formed under the
extended GEOM. This spectrum of allowed RVCs is consistent with the
URC, and given that the URC is constructed through the observations of
the velocity profiles of 1100 spiral galaxies, it is consistent with
observations as well. In fact, the two 
extreme RVCs predicted by the extended GEOM bracket the ensemble of
$V_{URC}$ shown in \cite{Salu2007}, while the median curve has a
form similar to both the $V_{URC}$ and $V_{NFW}$. Moreover, the
asymptotic behavior of URCs and $V_{NFW}$ is in good agreement with that of the RVC
with the most probable $p$ predicted by the extended GEOM. Importantly, we
have also shown that these stationary solutions in the galactic disk are stable under
perturbations, while in the galactic hub they are stable as long as
the period of oscillations of the perturbation is longer 
than $0.91_{\pm0.31}$ to $1.58_{\pm 0.46}$ billion years; these
perturbations have wavelengths shorter than $\sim 1.5 r_H^{*}$ to
$\sim 7r_H^{*}$. 

When comparing the graphs of the RVC obtained using the extended GEOM with the URC
in Fig.~\ref{Salu-Comp}, it becomes readily clear that
$v_{\infty}(x)$ may be too simplistic in the transition
region between the two asymptotic limits $x\to 0$ and
$x\to\infty$. This is  borne out by the 
shallowness of the minima in $S(q,p)$; we would expect that the more
accurate the choice of $v_{\infty}(x)$ is, the deeper the minima
will be. There are, in fact, many ways of smoothly joining together
the asymptotic behavior of the rotational
velocity profile in the two limits $x\to0$ and $x\to\infty$. Given
this, it is likely that future progress in 
determining the RVCs that can form under the extended GEOM using the
stationary-solution approach presented here will come from making a
different choice in $v_{\infty}(x)$ as much as, or even more than,
from more accurate numerical calculations. 

We have used a spherical model for our galaxy with the fluid rotating about
a single axis. Moreover, the values for the parameters $r_H^{*}$ and $v_H^{*}$
used here were obtained through observations of the motion of stars in spiral
galaxies. As such, the results we have obtained can be most directly
applied to the formation of spiral galaxies. It applicability to the
formation of other types of galaxies, such as those analysed in
\cite{Salu2012} and \cite{Paol2019}, is still an open question, and is
a topic of future research.

The extension of the GEOM we have considered here replaces the mass of
a test particle $m$ by $m\mathfrak{R}{c^2R/\Lambda_{DE}G}$ in the
Lagrangian for a test particle in general relativity. By doing so we
have changed the response of the motion of test particles to the
geometry of spacetime; the worldline of the test particles is now
determined by the extended GEOM, and not the GEOM. Einstein's field
equations are not changed, and the geometry of spacetime is still
determined by the solution of them. Importantly, this approach does not
differentiate between baryons and dark matter, and does not change the
worldlines of massless particles. It is for 
these reasons that we were able to show in \cite{ADS2010a} that the
extended GEOM is not excluded by the deflection of electromagnetic
waves by the Sun, or through the advancement of the perihelion of
Mercury. 

Another approach to modifying gravity, called modified gravity
theories in general, takes a different approach. The focus of these 
approaches is to change general relativity itself. Examples of such
theories include the Jordan-Brans-Dicke theory where the
gravitational constant is replaced by a scalar field, a scalar-tensor
theory where the cosmological constant is replaced by a scalar field,
and $f(r)$ theories where the Ricci scalar in the Hilbert action is
replaced by a function $f(R)$ of it (see \cite{Gan2019} for a
review). Such modifications of general relativity inherently changes
Einstein's field equations, and as such the resulting geometry of
spacetime. Importantly, the response of 
test particles to this geometry is not changed, and the worldline of
the particle is still determined by the GEOM. In particular, both the
worldlines of massive and massless particles are effected, and as
such Solar system tests of general relativity place stringent
limitations on such theories and the introduction of screening
mechanisms are needed (see \cite{Koya2016} for a review and
\cite{Lom2015} for an application of the screening). 

While the choice of $L_{\infty}(r)$, and the construction of
$v_{\infty}(x)$ was driven by observations and the requirement that
all the parameters used in $v_{\infty}(x)$ have a definite physical
interpretation, they are nevertheless choices. They were made with the
expectation that there are choices of initial conditions
$\mathbf{G}_0(r)$ which, when evolved to the stationary limit by
\textbf{\textit{Evol}}, will indeed result in both the $L_{\infty}(r)$
chosen and the stationary solution $\rho_{\infty}(r)$
resulting from this choice. Whether such expectation is born out is a
question that requires, if not solving the \textbf{\textit{Evol}},
then at the least a perturbative analysis of it near the stationary
limit. This analysis is also a focus of future research.

\section*{Declarations}

\paragraph{Funding and Conflicts of Interest:}
The author did
not receive support from any organization for the submitted work, nor
does the author have any financial or proprietary interests in any
material discussed in this article. 

\paragraph{Data Availability:} The data set containing the
results of the numerical calculations outlined in the paper, and used
to generate the figures that appear in it are available from the
author upon request.

\end{document}